\documentclass[runninghead]{llncs}
\pdfoutput=1

\usepackage{graphicx}
\usepackage{hyperref}
\usepackage{subfiles}
\usepackage{float}

\usepackage{mathtools}

\usepackage{lscape}
\usepackage[figuresright]{rotating}
\usepackage[graphicx]{realboxes}
\usepackage{adjustbox}
\usepackage{caption}
\usepackage{listings}

\usepackage{booktabs}
\usepackage{subfigure}
\usepackage{multirow}

\usepackage[ruled,linesnumbered]{algorithm2e}

\usepackage{tikz}
\usetikzlibrary{arrows.meta}
\usepackage{framed} 

\usepackage{amsmath}
\usepackage{amssymb}

\usepackage{lscape}
\usepackage{appendix}

\usepackage{comment}

\usepackage{setspace}
\usepackage{ntheorem}

\newtheorem{defi}{Definition}

\usepackage[
	n,
	operators,
	advantage,
	sets,
	adversary,
	landau,
	probability,
	notions,	
	logic,
	ff,
	mm,
	primitives,
	events,
	complexity,
	asymptotics,
	keys]{cryptocode}
	
\usepackage{dashbox}
\usepackage{tikz}
\usepackage{enumerate}

\usepackage{enumitem}

\begin{document}

\title{Security Analysis on Tangle-based Blockchain through Simulation}
\subtitle{(Full Version)\footnote{This paper (conference version) has been published at ACISP'20 \cite{qin2020security}.} }

\renewcommand\rightmark{}
\renewcommand\leftmark{Security Analysis on Tangle. B.Wang, Q. Wang.}
\newcommand{\romannum}[1]{\romannumeral #1}





\author{Bozhi Wang$^{\star\star}$\inst{1,3}, Qin Wang$\thanks{These authors contributed equally to the work.}$\inst{2,3}, Shiping Chen\inst{3}, Yang Xiang\inst{2} }
\authorrunning{B. Wang, Q. Wang}
\authorrunning{B. Wang, Q. Wang}
\titlerunning{Security Analysis on Tangle}

\institute{University of New South Wales
\email{(bozhi.wang@student.unsw.edu.au)} \\
\and
Swinburne University of Technology
\email{(qinwang,yixang@swin.edu.au)} \\
\and
CSIRO Data 61 
\email{(Shiping.Chen@data61.csiro.au)} \\
}

\maketitle           
\begin{abstract}
The Tangle-based structure becomes one of the most promising solutions when designing DAG-based blockchain systems. The approach improves the scalability by directly confirming multiple transactions in parallel instead of single blocks in linear. However, the performance gain may bring potential security risks. In this paper, we construct three types of attacks with comprehensive evaluations, namely parasite attack (PS), double spending attack (DS), and hybrid attack (HB). To achieve that, we deconstruct the Tangle-based projects (e.g. IOTA) and abstract the main components to rebuild a simple but flexible network for the simulation. Then, we informally define the three smallest actions to build up the attack strategies layer by layer. Based on that, we provide analyses to evaluate the different attacks in multiple dimensions. To the best of our knowledge, this is the first study to provide a comprehensive security analysis of Tangle-based blockchains. 

\end{abstract}


\section{Introduction}

Blockchain has emerged as the most prevailing technology in recent decades. In the early stage, blockchain was simply applied to the direct payment system, represented by Bitcoin \cite{nakamoto2008bitcoin}. Later, blockchain was employed as the distributed state machine to provide an executable environment for complicated tasks, represented by Ethereum \cite{wood2014ethereum} and EoS \cite{io2017eos}. However, more and more specific situations, such as IoT, micropayments, and edge computing, require high properties on scalability, performance, and cost. Therefore, Directed Acyclic Graph (DAG) is designed to improve the bottleneck of classic blockchain systems from the underlying structure. Different from the linear-chain in classic blockchains, DAG-based blockchain systems remove the limitation of blocks, expanding the network through a directed acyclic graph. Newly generated transactions without packing into blocks establish the network by directly confirming their parent transactions.   Through several iterative rounds, the main graph is formed with a low probability to be reversed.

Tangle structure is proposed by IOTA \cite{popov2016tangle}, one of the leading DAG-based systems. It aims to overcome the bottlenecks of current blockchain systems which include poor throughput on concurrency, low efficiency on performance, and high-cost on transaction fees. Tangle can be regarded as an expanding network formed by the continuously generated transactions. Transactions act as the smallest elements in the system to execute atomic operations, such as token transferring, witness validation, path extending, etc. The transaction-based Tangle possesses the properties of 1) \textit{High throughput}: Transactions can be attached to the network from different directions and verified by previous transactions in parallel without serious congestion. 2) \textit{High performance}: Newly arrived transactions are confirmed by the previous two transactions via a tiny Proof of Work (PoW) mechanism, where the computer consumption is negligible when compared to the traditional PoW. 3) \textit{Low cost}: There is no transaction fees in Tangle-based systems, which perfectly fits for the high frequent situations including IoT, micropayment, and edge computing.

However, Tangle-based blockchain systems confront the potential threats caused by the forks of subgraphs which spread in different directions. Following the specified tip selection mechanism \cite{popov2016tangle}, the system forms a multi-directional expansive network \cite{popov2017equilibria} to improve the scalability. More specifically, Tangle\footnote{In the rest of the paper, we use \textit{Tangle} to represent the Tangle-based blockchain systems. } achieves the delayed confirmation based on previously verified transactions in each direction, instead of an instant confirmation such as BFT-style consensus \cite{CastroL99}. The gap between delayed confirmation and instant confirmation leaves the blank of uncertainty and reversibility for attackers \cite{gervais2016security}, same as other probabilistic consensus like PoW \cite{nakamoto2008bitcoin}\cite{wood2014ethereum}. Existing chains are also threatened by miners who own insurmountable computing power to send massive transactions. Newly issued transactions are unpredictably attached to different subgraphs without control. Forks will frequently happen and leave the system under the risk of parasite chain attack and double-spending attack \cite{cullen2019distributed}. As a result, no leading subgraph can be formed to maintain the relatively stable states of the network.

In order to mitigate the issues, an additional centralized Coordinator is embedded in Tangle (e.g. IOTA project\footnote{\url{https://docs.iota.org/docs/the-tangle/0.1/concepts/the-coordinator?q=coodinator-\&highlights=coordinator\%27}}) to resist the potential attacks. Transactions snapshot by the Coordinator is immediately considered to be confirmed with 100$\%$ confidence. In addition, milestones are periodically broadcast by the Coordinator to reach the consensus across multiple subgraphs for stability and record the history by removing useless branches. Nodes accordingly rebuild the Merkle tree which contains the Coordinator's address to verify the milestone. This makes Tangle inherently a centralized system. IOTA official claims to cancel this centralized coordinator in the future. However, problems still exist without the Coordinator. The isolation and fork of each subgraph cannot be thoroughly solved. The network will inevitably spread into dispersed cliques under its inherent mechanism. We aim to provide comprehensive analyses of such risks by establishing three types of attack strategies. Our contributions are summarized as follows.

\begin{itemize}
    \item \textbf{Simulation of Tangle}: We deconstruct the Tangle-based system (IOTA) into main components including transaction generation, bundle unit, and selection algorithms. Based on that, we rebuild a simple but flexible simulation network. The simulation inherits the features of Tangle and provides an experimental environment for evaluations.
    
    \smallskip
    \item \textbf{Construction of attack strategy}: We define three actions as the basic benchmarks to construct our attack strategies layer by layer. The bottom layer (\textit{Layer0}) describes the role of each basic action. The media layer (\textit{Layer1}) presents the possible behaviors made up of actions. And the top layer (\textit{Layer2}) provides the attack strategies made up of combined behaviors. 
    
    \smallskip
    \item \textbf{Evaluations of attacks}: We evaluate the security through multiple metrics under different types of attack strategies. The discussions of potential influences include 1) Different proportions of behaviors at the same attack strategy; 2) Different attack strategies at the same parameter configuration; 3) Different parameter configurations at the same strategy.
    
\end{itemize}

The rest of the paper is organized as follows: The related works are provided in Section \ref{relatedwk}. The deconstruction of Tangle-based architecture and the rebuilding of the simulation are presented in Section \ref{deconstruct}. The constructions of our attack strategies are provided in Section \ref{construction}. Based on that, the implementation is shown in Section \ref{implementation}, followed by evaluations in Section \ref{evaluation}. Section \ref{sec-discuss} presents the discussions of simulated attacks. Finally, Section \ref{conclu} concludes our contributions and the future work.

\section{Related Work}
\label{relatedwk}

DAG \cite{dagwiki}, as a primitive in mathematics and computer science, is a finite directed graph with no directed cycles. Deeply rooted in the graph theory, DAG-based structure can be applied to various areas including data processing networks, genealogy and version history, citation graphs, data compression, etc. Currently, researchers and developers are trying to bring the DAG into the blockchain, to address the bottlenecks of scalability and performance. GHOST \cite{sompolinsky2013accelerating}\cite{sompolinsky2015secure}\cite{lewenberg2015inclusive}  as the backbone selection protocol provides an educational prototype for current DAG-based blockchain systems.  Unlike the classic Nakamoto consensus which chooses the longest chain, GHOST protocol selects the chain that holds the maximum sub-trees. It greatly improves throughput while holding the same block size. Inspired by GHOST, several blockchains replace the sub-tree structure into graph structure and redesign the whole consensus mechanism and network topology 
\cite{churyumov2016byteball}\cite{sompolinskyserialization}\cite{Nguyen2019StairDagCV}. Rather than focusing on the consensus at the block level, DAG prioritizes consensus at the transaction level in a separate mechanism. The transaction-based structure is inherently suitable for the micro-services, where Tangle-based blockchains exactly inherit the advantages. Although DAG is a competitive player, technical problems still exist. IOTA provides many strategies for the designation and protection, which are detailedly and explicitly described in \cite{popov2017equilibria}\cite{kusmierz2017first}\cite{popov2016tangle}\cite{kusmierz2018extracting}.

Several improvements by researchers are proposed to strengthen the potential weakness of Tangle. Cullen \textit{et al.} \cite{cullen2020resilience} proposes a matrix model to analyzes the efficacy of IOTA’s core MCMC algorithm, and present the improvement to resist parasite chain attacks. The matrix model clarifies the formulation for $\mathcal{H}$, to provide the explicit definition in the MCMC algorithm. Ferraro  \textit{et al.} \cite{ferraro2018iota} proposes a modified tip selection algorithm to make all honest transactions eventually be confirmed. The hybrid selection algorithm achieves the balance between biased preference for honest tips and high probability for older transactions. Gewu Bu  \textit{et al.} \cite{bu2019g} modifies the tip selection mechanism, called \textit{G-IOTA}, by choosing three verifying transactions at one time instead of two. Through the proposed algorithm, G-IOTA can tolerate several attacks.

Besides Tangle, there are also many other DAG-based systems. Byteball \cite{churyumov2016byteball} proposes the consensus based on a total order within DAG. The uni-direction is achieved by selecting a main chain to gravitate towards units issued by commonly recognized reputable witnesses. Spectre \cite{Sompolinsky2017SPECTRES} aims to establish the DAG structure based on concurrent and parallel block creation. It utilizes a recursive voting procedure where each block submits a vote for every pair of blocks. Accepted transactions are confirmed according to the votes. Hedera \cite{baird2016swirlds} develops the Hashgraph consensus algorithm as the underlying model. Inspired by BFT consensus, Hashgraph sets the 2/3 as the threshold, where a successful confirmation of newly generated transactions requires more than two-third witnesses from their ancestors. Nano \cite{nano} designs a low-latency cryptocurrency built on block-lattice data structure. Each account maintains an individual chain. Users update their chains asynchronously and keep track of their account balances rather than transaction amounts. All aforementioned DAG-based systems offer high scalability and low/no transaction fees. Graphchain \cite{boyen2016blockchain} proposes a DAG system following a similar tip selection rule with IOTA where one tip approves two ancestor transactions. The system additionally introduce a fee-based mechanism to establish the confidence of tips.

\section{Structure of Tangle Simulation}
\label{deconstruct}

Tangle is a permissionless network, providing an environment for data shared by all participants. In this section, we deconstruct the architecture of Tangle-based blockchains into three main components, which separately answer the question on how to generate a transaction, why need to bundle the transactions, and how to select the parent transactions for verification. Based on that, we abstract the key features of these fundamental components to rebuild a simple but flexible Tangle network for the simulation.

\medskip
\noindent{\textbf{Overview:}} Tangle bases on DAG where the vertex represents \textit{transaction} and the edge represents \textit{verification relationship}. Instead of the separation between making transactions by local users and achieving consensus by online miners, Tangle integrates these processes into one step. Whenever the transaction is generated and attached to Tangle, the consensus is simultaneously launched. Newly generated transactions are continuously attached to the network along with increased participants, which inevitably makes the subgraphs spread in different directions. In order to prevent the network split into isolated sub cliques, \textit{tip selection algorithm} is essential to lead the main graph in one direction to maintain the stability. Here, we provide a skeleton to show how Tangle forms and works, with the following procedures.

\begin{figure}[!t]
\centering
\includegraphics[width=1\linewidth]{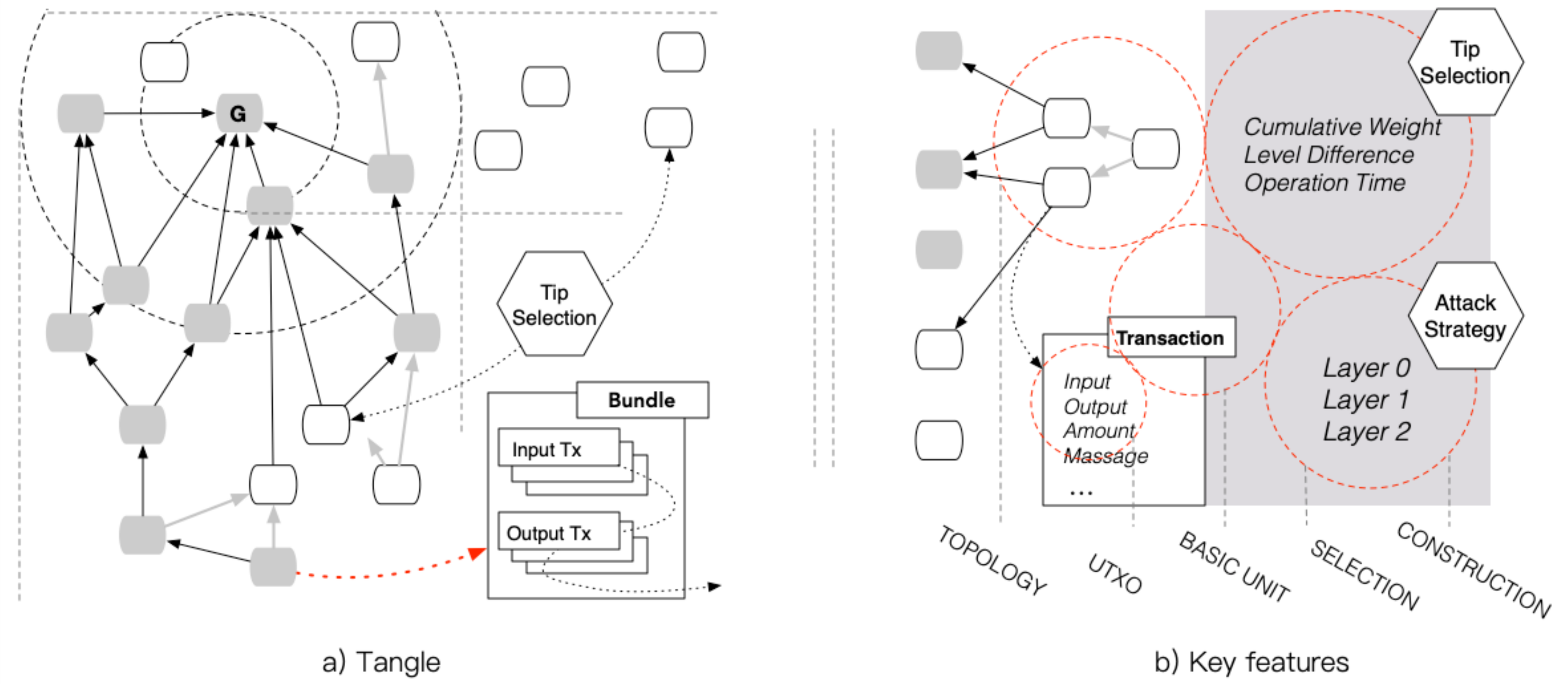}
\caption{Tangle with Its Features.}
\label{tangle}
\vspace{-10pt}
\end{figure}

\medskip
\noindent{\textbf{Generate a Transaction:}}
IOTA follows the model of UTXO, and transactions establish its network. Successfully generating a transaction requires to fill the fields of \textit{index}, \textit{address}, \textit{trunkTransaction} and \textit{branchTransaction}. The index is an accumulator, linearly growing along with the increased transactions. The address is used to identify transactions and generated by cryptographic sponge function. Subseed is an 81-tryte derived from the initialized seed and corresponding indexes under Keccak-384: \texttt{hash(seed + index)}. Each address in IOTA could only be spent once. If the address is used for the second time, it becomes poorly secure with high risk, due to the exposure of private subseed. Once a user withdraws the tokens from one address, it immediately creates a new address with an increased index. To achieve a complete round of withdraw/deposit, multiple pairs of private keys and addresses are necessary. As a consequence, it leads to the concept of the bundle.

\textit{Insight:} The transactions are the smallest components serving for further construction in Tangle. Every operation of token transferring and transaction verification is based on newly generated transactions. The trends of the network are caused by the behaviors of transactions. Therefore, we focus on the transactions and analyze potential behaviors that may frequently happen. The transaction can be used as an approver for both honest behaviors and malicious behaviors of the previous transactions. We define three basic actions to build up the behaviors surrounding a transaction in \underline{Section \ref{construction}}.

\smallskip
\noindent{\textbf{Packaged Transactions as Bundle:}}
Bundle, the basic unit structure in IOTA, is a top-level construction that links related transactions into one clique. The bundle itself cannot be broadcast, instead, a collection of individual transactions are broadcast. All transactions can be regarded as part of the bundle, and the metadata is recorded on every single transaction (in \textit{trytes} format) instead of the virtual bundle. In other words, a bundle can be reconstructed from the transaction collection at any time through the fields of \textit{bundle hash}, \textit{index}, \textit{trunkTransaction} and \textit{branchTransaction}. The trunkTransaction field is used to link transactions in the same bundle where a transaction with a higher index can trace back to lower index by targeting their unique trunkTransaction strings. The branchTransaction field is used to connect bundles in Tangle. All transactions in the same bundle are filled with the same branchTransaction string, except for the initial transaction and tail transaction. These two special transactions distinguish one bundle from another, bridging different units in the network. Details are shown at \underline{Fig.\ref{bundle} in Appendix D}.

\textit{Insight:} The bundle is the basic unit serving for money transferring. There are no real entities in the network, and all the steps are executed through transactions. Collective transactions inherently share the same seeds from one node, so that we regard the bundle as a separate vertex in the network. Packaging collective transactions into one bundle guarantee the security level in an open environment. In our simulation, we ignore the bundle structure and only abstract the key features of the bundle, (such as PoW, parent selection, etc.).  We apply these features into transactions for simplicity in a closed testing environment, as shown in \underline{Section \ref{construction}.}

\medskip
\noindent{\textbf{Tip Selection Algorithm:}}
Tip\footnote{Tip means the newly generated transactions that have not been validated.} selection represents the selection strategies of newly generated transactions. The strategy of Tangle follows the principle: one tip selects and approves two ancestor transactions. This principle decides the direction of the graph as shown in \underline{Fig.\ref{tangle}.a}. As above-mentioned, transactions in Tangle are organized in bundles, where the tail transaction is selected by the approver through the trunkTransaction field for final consistency. The trunkTransaction field connects different transactions in the same bundle, shown in the green lines in \underline{Fig.\ref{bundle}}. Each transaction with a higher index can trace back to a lower index. The branchTransaction field inside the bundle is slightly different, all of the transactions are filled with the same string, except for the field in initial and tail transactions. The initial and tail transactions are generated by tip selection mechanism to connect different bundles, as the blue lines in \underline{Fig.\ref{bundle}}.

There are three kinds of Tip selection mechanisms provided in \cite{popov2016tangle}:  \textit{Uniform Random}, \textit{Unweighted Random Walk} and \textit{Weighted Random Walk}. Note that the Markov Chain Monte Carlo Algorithm (MCMC) is based on a weighted random walk. For the higher probability of being selected, several matrices are proposed for better evaluation. \textit{height (h)} represents the length of the longest path, equal to the path from genesis to the current transaction. \textit{depth (d)} means the longest reverse-oriented path, equal to the path from current transaction to a certain tip. \textit{Cumulative weight ($\mathcal{W}$)} is counted as the number of transaction \textit{v} being verified both directly and indirectly. $\mathcal{W}$ reflects the probability of a tip being selected by the random walk algorithm. $\alpha$ is the configurable parameter to control the effectiveness of $\mathcal{W}$ in MCMC. When $\alpha$ converge towards 0, tip selection becomes uniformly random, while towards 1, tip selection becomes deterministic.

\textit{Insight:} Tip selection describes how a newly generated transaction selects its ancestors. No matter which detailed mechanism the node chooses, the selection processes are all based on the variants of the random algorithm. The algorithm guarantees a tip can select in the large range without strict rules. We divide the selection algorithm in a general way shown in \underline{Section \ref{construction}}.

\medskip
\noindent{\textbf{Mapping to Simulation:}} Based on the deconstruction of Tangle, we conclude three insights as to the guidelines for our attack simulations of Tangle. Then, we capture the key features as in \underline{Fig.\ref{tangle}.b} on how to rebuild our simulation networks. Here are the details.

\begin{itemize}
    \item For the structure, the simulation is based on the UTXO model proposed by Bitcoin \cite{nakamoto2008bitcoin}. The UTXO model is naturally fit for DAG-based blockchain systems since there is no account in DAG. The UTXO model is responsible for the correctness of the balance.
    
    \item For the basic unit, the simulation employs the transaction directly as the smallest unit instead of the bundle. In a closed environment, the key-exposure problem can be ignored. The transaction-based exchange provides us a flexible reconstruction.
    
    \item For the topology, the simulation follows the original pattern of Tangle: every tip verifies two other parent transactions by selection mechanism. The selection mechanism is determined by factors as discussed in \underline{Section 5.2}. 
    
    \item For the tip selection mechanism, we capture three most influential factors in Tangle including \textit{cumulative weight}, \textit{level difference}, and \textit{operation time}. Tips select their parent transactions according to the \underline{Equation \ref{decision-parenttip}}.
     
    \item For the attack strategy construction, we build three typical attacks confronted by Tangle, containing PS, DS, and HB. The constructions of attack strategies are built on the smallest actions layer by layer.
\end{itemize}

\section{Construction of Attack Strategies}
\label{construction}

In this section, we provide three main attack strategies: Parasite Attack (PS), Double Spending Attack (DS), and Hybrid Attack (HB). Each strategy represents a family of concrete attacks that inherit the same foundations. To make it clear, we start from the smallest units to progressively establish a practical attack. We construct the strategies by three layers: \textit{layer0} for unit actions, \textit{layer1} for atomic behaviors and \textit{layer2} for combined attack strategies. Note that, the actions and behaviors defined below are mainly focused on the malicious nodes since honest nodes will conduct honest transactions and behaviors. Only malicious/dishonest nodes have multiple combinations. Here, we provide the details.

\subsection{Layer0: Unit Actions}

We define the smallest unit actions in the bottom layer, denoted as \textit{layer0}, to describe the activities and behaviors a node can select. It can be regarded as a binary selection at each unit action, and the combination of unit actions make up an atomic behavior in \textit{layer1}. More specifically, we list three-unit actions as the metrics, including:

\begin{itemize}
    \item \textit{Action A:} The unit action A represents whether a newly generated transaction is 1) \textit{valid} [$A_1$] or 2) \textit{invalid} [$A_2$]. 
    \item \textit{Action B:} The unit action B represents whether a newly generated transaction is attach to the parent tips 1) by the  \textit{random selection} mechanism [$B_1$] or 2) by selecting the transactions which are issued by same nodes/entities with itself [$B_2$]. Here we call $B_2$ as  \textit{selfish selection}.
    \item \textit{Action C:} The unit action C represents whether a newly generated transaction is selected from 1) the pool with valid transactions [$C_1$] or 2) the pool with invalid transactions [$C_2$]. We denote the first one as the \textit{valid pool} and the second as the \textit{invalid pool}.
\end{itemize}

We can see that each set of unit action are made up by two possible choices to describe the instant state. We denote this process as a \textit{binary selection} for simplicity and clarity. We use $1$ to represent the selections of elements in the first line $\{A_1, B_1, C_1\}$, while we use $0$ to represent the selections of elements in the second line including $\{A_2, B_2, C_2\}$. We summarize each possible selection in \underline{Table \ref{unit}}. It should be noted that \textit{selfish} selection means a newly generated transaction and the parent transaction is issued by the node with the same identity (either honest or malicious).

\begin{table}[!hbt]
 \caption{Unit Action}
  \centering
  
    \begin{tabular}[t]{lll}
    \toprule
    \textit{Action A} & \textit{Action B}  & \textit{Action C}\\  
    \midrule
    Valid Tx [$A_1$] & Random Selection [$B_1$] & Valid Pool [$C_1$] \\  
    Invalid Tx [$A_2$] & Selfish Selection [$B_2$] & Invalid Pool [$C_2$]  \\  
    \bottomrule
    \end{tabular}
    \label{unit}
\end{table}

\subsection{Layer1: Atomic Behaviors}

\textit{Layer1} is a collective set of various behaviors made up by different combinations of unit actions. The behavior assists to describe how to generate a transaction at the initial stage. The behaviors are atomic for the construction of attack strategies and are also closely related to the category of attack types. Here, we summarize feasible atomic behaviors in \underline{Table \ref{behavior}}. Note that, every single behavior covers the unit actions A, B, and C. We employ the binary selection $0$ and $1$ to distinguish the combinations of unit actions.

\begin{table}[!hbt]
 \caption{Atomic Behaviors}
  \centering
    \begin{tabular}[t]{cccc}
    \toprule
    \textit{Binary Selection} \,\,& \textit{Action Combination} \,\, & \textit{Feasibility} \,\,& \textit{Index} \\  
    \midrule
    111 & ($A_1,B_1,C_1$) & Y & a \\  
    110 & ($A_1,B_1,C_2$) & Y & b \\  
    101 & ($A_1,B_2,C_1$) & Y & c \\
    100 & ($A_1,B_2,C_2$) & N & - \\
    011 & ($A_2,B_1,C_1$) & Y & d \\
    010 & ($A_2,B_1,C_2$) & Y & e \\
    001 & ($A_2,B_2,C_1$) & Y & f \\
    000 & ($A_2,B_2,C_2$) & N & - \\
    \bottomrule
    \end{tabular}
    \label{behavior}
\end{table}

From Table \ref{behavior}, we can see that there are $8$ possible \textit{atomic behaviors}  which are used to show how a malicious node executes the transactions. Take $101$ - ($A_1, B_2, C_1$) as an example, it means the behavior that \textit{A malicious node generates a valid transaction, being selfishly attached to the parent tips from the valid pool.} This behavior is feasible in the simulation without logic error, denoted as $Y$ in the $Feasibility$ column. On the contrary, the behavior $100$ is infeasible, since no invalid transaction exists in the network if the malicious nodes only send valid transactions. Similarly, the behavior $000$ is infeasible since for a malicious node. Selfishly attaching processes from the invalid pool is equal to the random selection from the invalid pool, which means $000$ is equal to $010$. Therefore, only $6$ of the behaviors are feasible, and we mark them with indexes from $a$ to $f$.

\subsection{Layer2: Combined Attack Strategies}

Based on the behaviors in $layer1$, we construct attack strategies in $layer2$. We identify three types of attacks, including Parasite Attack (PS), Double Spending Attacks (DS) and Hybrid Attacks (HB). The parasite attack means an attacker secretly creates a sub-Tangle with high weights and attach this sub-graph at some time to the network. It may reverse the main Tangle when newly attached transactions are widely distributed. Double spending attack means an attacker creates the same transaction more than one time. This attack splits Tangle into two branches so that s/he can spend one coin multiple times in different branches.

PS and DS are pure attacks without the overlap of behaviors, whereas HB represents the attacks containing overlapping behaviors, such as $f$ (Detailed explanation in the later paragraph). Here, we provide the decision principle, denoted as $\Phi$, on how to categorize the attacking types. We conclude four stages of the decision processes shown in \underline{Equation \ref{decision-a}}.

\begin{equation}
\left.
\begin{array}{ll}
\begin{aligned}
    \textrm{Confusion behavior} - stage 1 \\
    \textrm{Send the invalid Tx} - stage 2   \\
    \textrm{Verify the invalid Tx} - stage 3 \\
    \textrm{Overlapping behavior} - stage 4  \\
\end{aligned}
\end{array}
\right]    \Rightarrow \Phi
\label{decision-a}
\end{equation}

For the $stage 1$, the confusion behavior means the malicious node pretends to act as an honest node, such as the behavior $a$: \textit{A malicious node sends a valid transaction, being randomly attached to the parent tips from the valid pool.} We cannot obtain any useful knowledge to distinguish whether the transaction is issued by an honest or malicious node. $stage 2$ represents a malicious node sends invalid transactions, where related behaviors in $layer2$ are $(d,e,f)$. $stage 3$ refers to the parent tip selection, and related behaviors are $(b,e)$. $stage 4$ provides the overlapping behaviors including  $(c,f)$. Therefore, the general decision principle $\Phi$ is shown in \underline{Equation \ref{decision-b}}. The symbol ``-'' means do not exist.

\begin{equation}
 \, (a,-) \,|\, (d,e,f,-) \,|\, (b,e,-) \,|\, (c,f,-) 
\label{decision-b}
\end{equation}

Alternatively, based on the $\Phi$, we provide the concrete decision principle for the PS, DS, and HB, which are separately denoted as  $\Phi$[PS], $\Phi$[DS] and  $\Phi$[HB] as in the \underline{Equation \ref{decision-c}}. The key principle of PS is to selfishly select parent transactions, while the principle of DS is to send/verify invalid transactions. But there are some overlapping behaviors and logic errors in the strategy. Therefore, we provide each attacking type by specified decision principles. The detailed principle is listed in the \underline{second column of Table \ref{attack}}.

\begin{equation}
 \{\, \Phi \,\,|\,\,  \Phi\textrm{[PS]},\,\,\Phi\textrm{[DS]},\,\Phi\textrm{[HB]}\, \}
\label{decision-c}
\end{equation}

Based on the decision principles for each attacking type, we back to the reason why the behavior $f$ is an overlapping behavior. The behavior $f$ means: \textit{A malicious node sends an invalid transaction, being selfishly attached to the parent tips from the valid pool}. On the one hand, $f$ selfishly selects its parent tips, satisfying the condition of PS ($\Phi$[PS]). On the other side,  $f$ sends an invalid transaction to the network, satisfying the condition of DS ($\Phi$[DS]). Therefore,  $f$ is an overlapping behaviors applied in the hybrid attacks ($\Phi$[HB]).

\begin{table}[htb!]
 \caption{The Performance of Increased User Size}
 \label{attack}
\resizebox{\textwidth}{!}{
\newcommand{\tabincell}[2]{\begin{tabular}{@{}#1@{}}#2\end{tabular}}
  \centering
   \begin{tabular}[t]{cccc}
        \toprule
            \tabincell{c}{\textbf{Attack} \\ \textbf{Types} \\ } &
            \tabincell{c}{Decision\\ Principle } &
            \tabincell{c}{Feasible \\ Behavior}&
            \tabincell{c}{Attack \\ Strategies} 
            \\
        \midrule
        & (a,-) $|$  (d,e,f,-)  $|$  (b,e,-)  $|$   (c,f,-)    &  & \\
        \cmidrule{2-2} 
        PS& (a, -) $|$ - $|$ - $|$ (c)    & c,f & c,ac (2)\\ 
        \cmidrule{4-4} 
        \multirow{2}{*}{DS} & \multirow{2}{*}{(a,-) $|$ (d,e) $|$ (b,e) $|$ - }   & \multirow{2}{*}{b,d,e,f } & e,ae,bd,de,abd,\\
              &   &     & ade,bde,abde (7)  \\  
        \cmidrule{4-4} 
        \multirow{3}{*}{HB} & \multirow{3}{*}{(a,-) $|$ (d,e,f) $|$ (b,e) $|$ (c,f)}    & \multirow{3}{*}{f} & ce,bf,ef,cef,bcf,bef,bce,def,cde,\\
            &   &     &  bdf,bcd,aef,acef,abf,abcf,ace,  \\  
                &      &  &  abef,abce,adef,acde,abdf,abcd (22)  \\  
        \bottomrule
    \end{tabular}}
\end{table}

\section{Implementations}
\label{implementation}

\subsection{Parameters and Notations}

In this subsection, we provide the notations used in our implementations and testings. There are two types of parameters in the system, the first are binary parameters such as $(A_1, A_2)$ as discussed in \underline{Section \ref{construction}}, used for the construction of attacks. Since the transactions are atomic in the simulation, we ignore duplicated introductions. The second are continuous parameters such as operating time, number of total transactions, etc. They are used for adjusting the configurations during the simulation. The related continuous parameters are denoted below.

The parameters include total transactions $T$,  honest transaction $H$, invalid Transaction $F$, intervals between two newly generated transaction $D$, time of PoW $I$, height of block $h$, simulation operating time $\mathcal{T}$, level difference $\mathcal{L}$ and cumulative weight $\mathcal{W}$. Besides, derived parameters also provided: strategy space $\mathbb{S}$, transaction generation speed $T/(D+I)$, the ratio of invalid transaction and the total transaction $ratio(\mathcal{F})=F/T$, and the ratio between different behaviors in one strategy $ratio(\mathcal{B})=x:y:z$ (short for $xyz$), where $xyz$ depends on initial settings when launching the attacks.\\

\subsection{Key Principles}

The growth of a DAG is based on the continuously generated transactions. Previous transactions get weighted when tips are attached, measured by \textit{cumulative weight ($\mathcal{W}$)}. Two aspects are considered: the configuration of each unit weight and the methods to select parent transactions. For the first side, the unit weight of every single transaction randomly varies from $1$ to $4$, to provide a better simulation for the real scenario. For the second side, a parent selection mechanism is required to take key metrics into consideration which contains the cumulative weight, operation time, level difference (level represents the transactions with the same height). Here, we present more explanations on these three metrics. 

\begin{itemize}
    \item \textit{Level Difference:} Denoted as $\mathcal{L}$, level represents the transactions that are identified with the same heights.  Level Difference is the distance of the heights between current tips and the selected parent transactions. 
    \item \textit{Cumulative Weight:} Denoted as $\mathcal{W}$, the cumulative weight is calculated as an accumulated value when tips are attached. It is the sum of each unit weight of the attached tips. The unit weight randomly varies from 1 to 4 to avoid fraud.
    \item \textit{Operation Time:} Denoted as $\mathcal{T}$, the operation time represents the executed time of a transaction since it was generated. A transaction will be discarded when times out.
\end{itemize}

We provide the following principles for tip selection and abandon of invalid transactions in the implementation. Tip selection decides how tips choose their parent transactions. Invalid transaction decision shows how a transaction is discarded. These principles limit the size of networks under control.

\begin{defi}[Mechanism For Tip Selection]
\begin{equation*}
\begin{array}{ll}
\begin{aligned}
    p =  3 \times | 15  - \mathcal{W}_i| &+\frac{100}{5^{\mathcal{L}}}- 1.5^{\frac{\mathcal{T}}{60}} \quad (p\geq 0)\\
   where, \,\,\,w[1] &= w_c \quad (\mathcal{L}=1)  \\
    w[i]& =  0.8 w[i-1] \quad (\mathcal{L}=2-6)   \\
  	& = 0.9 w[i-1] \quad 	(\mathcal{L}=7-16) \\
    & = 0.01 w[1]  \quad (\mathcal{L}=17-29)  \\
    \mathcal{W}_{i} & = \mathcal{W}_{i-1} + w[i],\quad where,\,\,  i\in \mathcal{L} \\
\end{aligned}
\end{array}
\end{equation*}
\label{decision-parenttip}
\noindent where $\mathcal{L}$ is height difference, $\mathcal{W}$ is cumulative weight, $w[i]$ represents individual round weight, and $w_c$ is current weight. $p$ represents tip selection probability.
\end{defi}

From \underline{Definition \ref{decision-parenttip}}, we can see that the selection probability $p$ is mainly influenced by three factors:  $\mathcal{L}$,  $\mathcal{W}$ and $\mathcal{T}$. $\mathcal{L}$ varies inversely with the $p$, which means a tip tends to be impossibly selected as the level difference increases.  $\mathcal{W}$ is an iterative algorithm within three intervals according to $\mathcal{L}$. The equation at different intervals has a different decay rate. The tips are required to be smoothly decayed with a small level difference (near to the latest transaction) while be sharply decayed at the high difference. $\mathcal{T}$ is used to prevent the tip from being suspended for too long. We provide a slice of example codes in \underline{Appendix C} for a better explanation.

\begin{defi}[Decision For Invalid Transaction]
A newly generated transaction will be discarded, as \texttt{Tx} $=\bot$, when triggering the conditions:
\[ \{ \texttt{Tx}=\bot\,|\underline{\,\mathcal{L} >30 \, || \,\mathcal{W}<30}  \,\,  \cap   \,\, \underline{\mathcal{T}>1000s }\}\]
\noindent where $\mathcal{L}$ is the height difference, $\mathcal{W}$ is the cumulative weight. $\mathcal{T}$ represents the operating time.
\label{decision-invalid}
\end{defi}

From \underline{Definition \ref{decision-invalid}}, we obtain that a newly generated transaction will be deemed as invalid when exceeding the thresholds either on the specified level difference (30) or on operation time (1000 s). Our simulations and evaluations of attack strategies only consider valid transactions.

\subsection{Implementation Logic}

Our implementation is based on the simulation of Tangle and the construction of attack strategies. We provide detailed workflows including steps on receiving the transactions from peer nodes, generating/sending new transactions, and launching the attack strategies. Detailed example codes are referenced in Github\footnote{Source Code: \url{https://github.com/BozhiWang/Tangle-based-Blockchain-attack-simulation}.}.


\smallskip
\noindent \textbf{Logic.1. Launch the Attack Strategy:}
\begin{itemize}
    \item \textit{step1:} Configure all the initial parameters including total transactions, Ratio($\mathcal{F}$), Ratio($\mathcal{B}$), and operating time $\mathcal{T}$. 
    \item \textit{step2:} Select the attack types based on strategies (see the \textit{layer2}) from the behaviors (see the \textit{layer1}) and the actions (see the \textit{layer0}).
    \item \textit{step3:} Set different parameters of Ratio($\mathcal{F}$) and Ratio($\mathcal{B}$) for the test goals (see in Section \ref{sec:goals}).
    \item \textit{step4:} Launch the atttack transactions by \textit{Send New Transactions} and \textit{Receive New Transactions}.
    \item \textit{step5:}  Collect the results for evaluation and analysis. (see \underline{Section \ref{evaluation}}).
\end{itemize}

\smallskip
\noindent  \textbf{Logic.2 Receive the Transactions:}
\begin{itemize}
    \item \textit{step1:} Listen to the peer nodes to receive the transactions, with up to $n$ transactions per second.
    \item \textit{step2:}  Put the valid transactions into the valid pool and the invalid transaction into the invalid pool for the verification.
    \item \textit{step3:}  Calculate the maximum height of the current DAG.
    \item \textit{step4:}  Count the cumulative weights $\mathcal{W}$ for the parent transactions through weight iteration in Definition \ref{decision-parenttip}.
    \item \textit{step5:} Remove the timeout/expired transactions from the transaction pool according to the \underline{Definition \ref{decision-invalid}} and change the current status of transactions.
\end{itemize}

\smallskip
\noindent  \textbf{Logic.3. Send New Transactions:}
\begin{itemize}
    \item \textit{step1:} Generate transactions with fixed parameters including height, weight, timestamp.
    \item \textit{step2:} Select two parent transactions to verify according to the tip selection mechanism.
    \item \textit{step3:} Launch the PoW verification for attach tips.
    \item \textit{step4:} Broadcast the transaction for times.
\end{itemize}

\subsection{Implementation Goals}
\label{sec:goals}

We provide four goals to evaluate attack strategies under different configurations. Goal I aims to test different types of attacks strategies. Goal II mainly tests combined behaviours. Goal III is to test the influence of total nodes. Goal IV is focused on the selfish strategy. Detailed parameter settings are presented at  \underline{Table \ref{tab-goals} in Appendix B}.

\smallskip
\textbf{Goal I:}
The first goal aims to test the influence of different attacks strategies (mainly hybrid attacks), namely ($\mathbb{S}$). The initial configurations include the attack strategies of each type and the rate of malicious nodes. Specifically, we set the total nodes as 100, and configure the initial rate of malicious node to be 20$\%$. The main variables are focused on the different strategies of attacks ($bd,be,ace,abe,ade,e,abcd,abdf$) with corresponding Ratio($\mathcal{B}$) on each strategy. The Ratio($\mathcal{B}$) is set to be $\{55,433,4222\}$\footnote{We employ the abbreviation ``$433$'' to represent the ratio of ``$4:3:3$'' for simplicity which is equally applied to other ratios. } respectively, for the strategies containing two, three, and four behaviors. Detailed variable configurations are shown in the testing sets of (\textit{Set1, Set2, Set3}).

\smallskip
\textbf{Goal II:}
The second goal is going to test the influence of different ratios of combined behaviors, namely Ratio($\mathcal{B}$). The initial configurations include three randomly selected attack strategies in PS/DS ($abe, ade abd$). The total nodes are set to be 100 with 10$\%$ malicious nodes. The main variables include different ratios of the combined behaviors, where Ratio($\mathcal{B}$)$=\{811,622,433,631,613\}$. We provide three testing sets with different Ratio($\mathcal{F}$) on $\{10\%,20\%,30\%\}$. Detailed configurations are listed in the testing sets of (\textit{Set4, Set5, Set6}).

\smallskip
\textbf{Goal III:}
The third goal is aimed to test the influence of total nodes. The initial configurations include the random selected attack strategies ($abe, ade abd$) with the ratios of combined behaviors $622$. The variables are the total nodes $T$ where $T=\{20,50,100\}$. We present the testing sets in different ratios of malicious nodes where Ratio($\mathcal{F}$)=$\{10\%,20\%,30\%\}$. Detailed configurations are provided in the testing sets of (\textit{Set7, Set8, Set9}).

\smallskip
\textbf{Goal IV:}
The fourth goal is focused on the selfish strategy combined by $ac$. The testing contains the influence by the variable Ratio($\mathcal{B}$) in $Set10$, the different total nodes in $Set11$ and the changeable Ratio($\mathcal{F}$) in $Set12$. Also, the testing items can be referenced and compared across these three sets, since the targeted strategy $ac$ is fixed. Detailed variable configurations are in presented in testing sets of (\textit{Set10, Set11, Set12}).

\section{Evaluation Analyses}
\label{evaluation}

Based on Goals, our experiments provide different types of results. In this section, we firstly give the specified inputs and outputs, and then show the trends of different testing sets. Detailed data and other outputs are presented in \underline{Appendix D}.


\subsection{Analysis on Result I}
In the simulation I, we set totally eight attack strategies $\{ \mathbb{S} \,|\, \textit{bd},\textit{be},\textit{ac},\textit{abe}, \textit{ade}, \textit{abd}, \\ \textit{e},\textit{abcd},\textit{abdf}\}$, the corresponding Ratio($\mathcal{B}$) of each strategy, and three sets of Ratio($\mathcal{F}$)=$\{ \mathcal{F} \,|\, 10\%\}$ as the input parameters. The outputs contain the confirmed invalid transactions, the confirm time, the abandoned invalid transactions and the abandoned valid transactions. From the \underline{\textit{Result I} in Figure \ref{resulti}}, we can find the trend caused by different factors. (1) For the same strategies, no matter how they are made up, such as \textit{ade} \textit{e} and \textit{abcd}, the confirmed invalid transactions are increasing along with the number of malicious nodes in a positive correlation. The confirm time varies in a range of 200-800s. The abandoned transactions significantly increase with the number of malicious nodes.  (2) For the different hybrid strategies, Ratio($\mathcal{F}$) has different influences on them. Several strategies are sensitive to the changes like $abe,be$. (3) The abandoned invalid transactions and valid transactions increase at the same time along with changes on Ratio($\mathcal{F}$) where malicious nodes have a significant influence.

\subsection{Analysis on Result II}

In the simulation II, we set three sets of strategies $\{ \mathbb{S} \,|\, \textit{abe}, \textit{ade}, \textit{abd}\}$, three  Ratio($\mathcal{F}$)=$\{ \mathcal{F} \,|\, 10\%, 20\%, 30\%\}$, and five Ratio($\mathcal{B}$)= $\{ \mathcal{B} \,|\, 811, 622, 433, 631, 613\}$ as the input parameters. The outputs are the ratios between invalid transactions and total transactions.  From the \underline{\textit{Result II} in Figure \ref{resultii}}, we can find that (1) For the Ratio($\mathcal{F}$) on the same strategies, such as \textit{ade} (the blue column in the histogram), invalid transactions will significantly increase along with the malicious nodes. The trend is determinate for such situations. (2) For the different strategies, we can find that the Ratio($\mathcal{F}$) has different influences on them. $ade$ varies monotonously with the ratio, while the other two have a peak value at a certain ratio. (3) The attack is sensitive to some behaviors such as $b$. Invalid transactions in strategies containing $b$ ($abe,abd$) are more significant than the strategies without $b$.

\subsection{Analysis on Result III}

In the simulation III, we set three attack strategies $\{ \mathbb{S}\, |\, \textit{abe}, \textit{ade}, \textit{abd}\}$ as the basic testing strategy with initial Ratio($\mathcal{B}$)=$\{ \mathcal{B} \,| \, 622\}$. There are three  Ratio($\mathcal{F}$)= $\{ \mathcal{F} \,|\, 10\%, 20\%, 30\%\}$ and four sets of total nodes $\{Tx\,|\, 20, 50, 100,200 \}$ as the input parameters. The outputs are the ratio between confirmed invalid transactions and total transactions. The ratio provides a direct and visualized relationship. From the \underline{\textit{Result III} in Figure \ref{resultiii}}, we can find (1) For the same strategy, such as \textit{ade} (the blue column in the histogram), the trend of ratio is relatively stable under different Ratio($\mathcal{F}$). (2) The ratio maintains stable when the Ratio($\mathcal{B}$) increases. This also means the ratio of invalid transactions with total transactions varies slightly with the malicious nodes. The number of malicious nodes has little influence on the ratio. (3) For the different combinations, the strategies like $ade$ are more sensitive to variations than the strategies like $abe,abd$. The results show significant differences in these strategies.

\subsection{Analysis on Result IV}

\begin{sidewaysfigure}[p]
\centering
\subfigure[Result I]{
\begin{minipage}[b]{0.23\linewidth}
\includegraphics[width=1\linewidth]{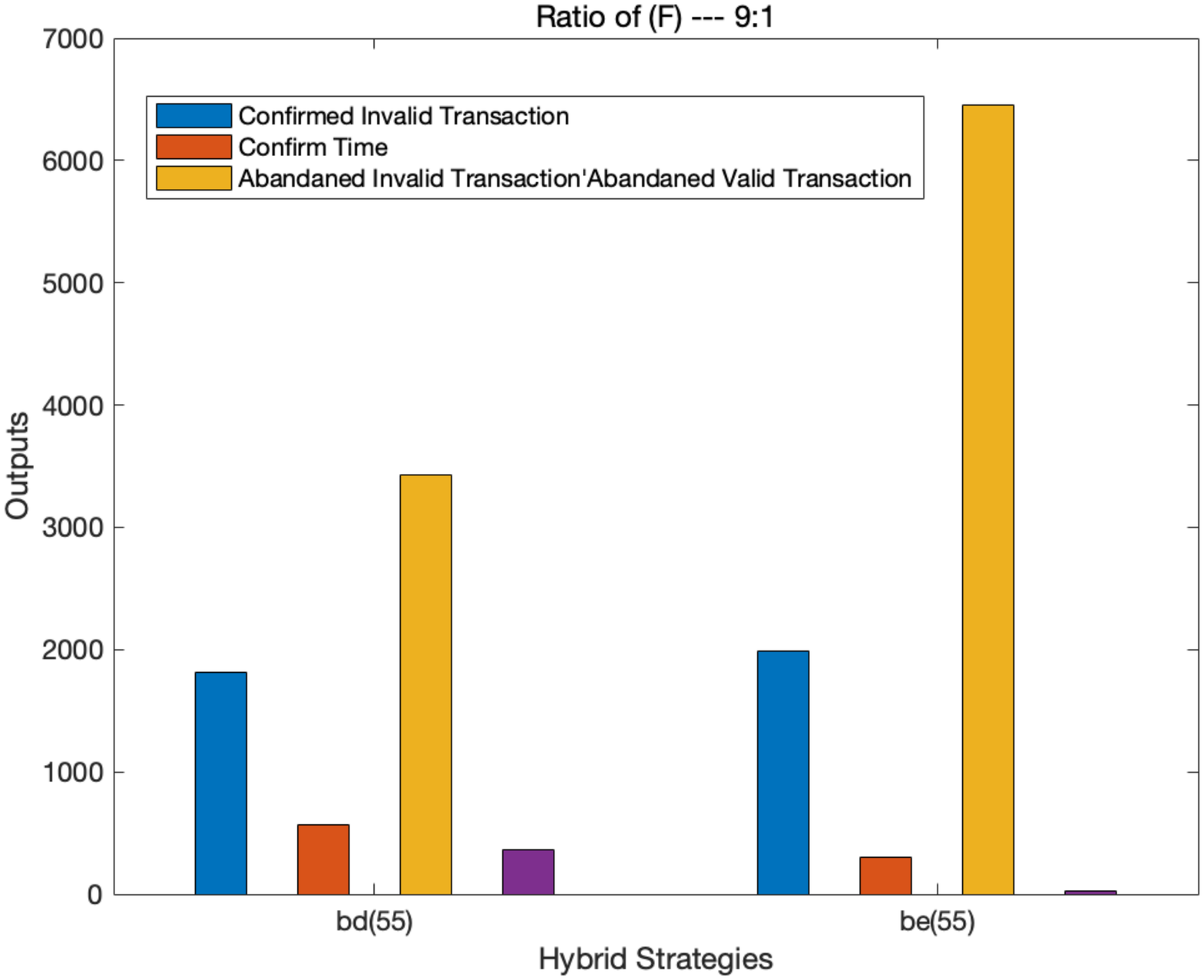}\vspace{4pt}
\includegraphics[width=1\linewidth]{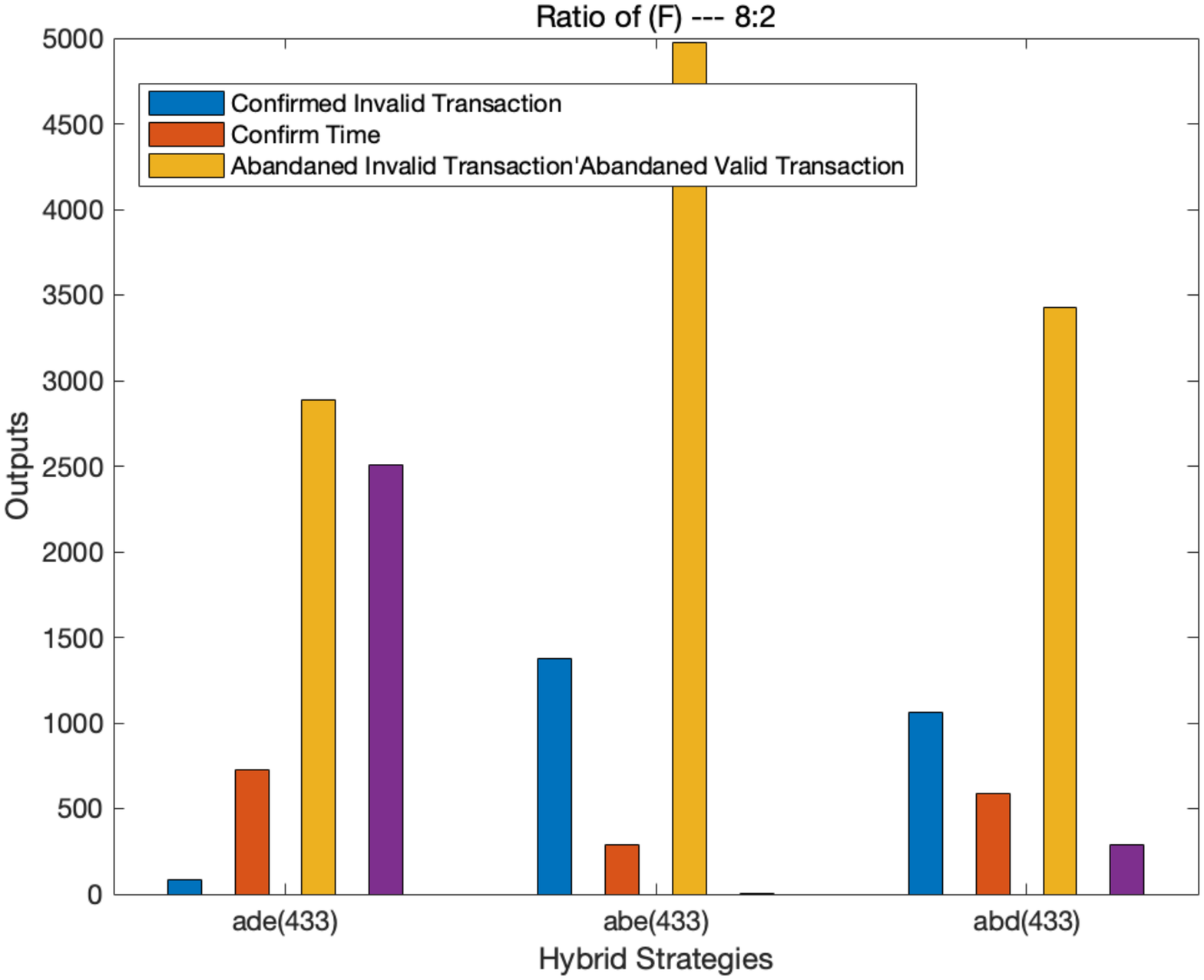}\vspace{4pt}
\includegraphics[width=1\linewidth]{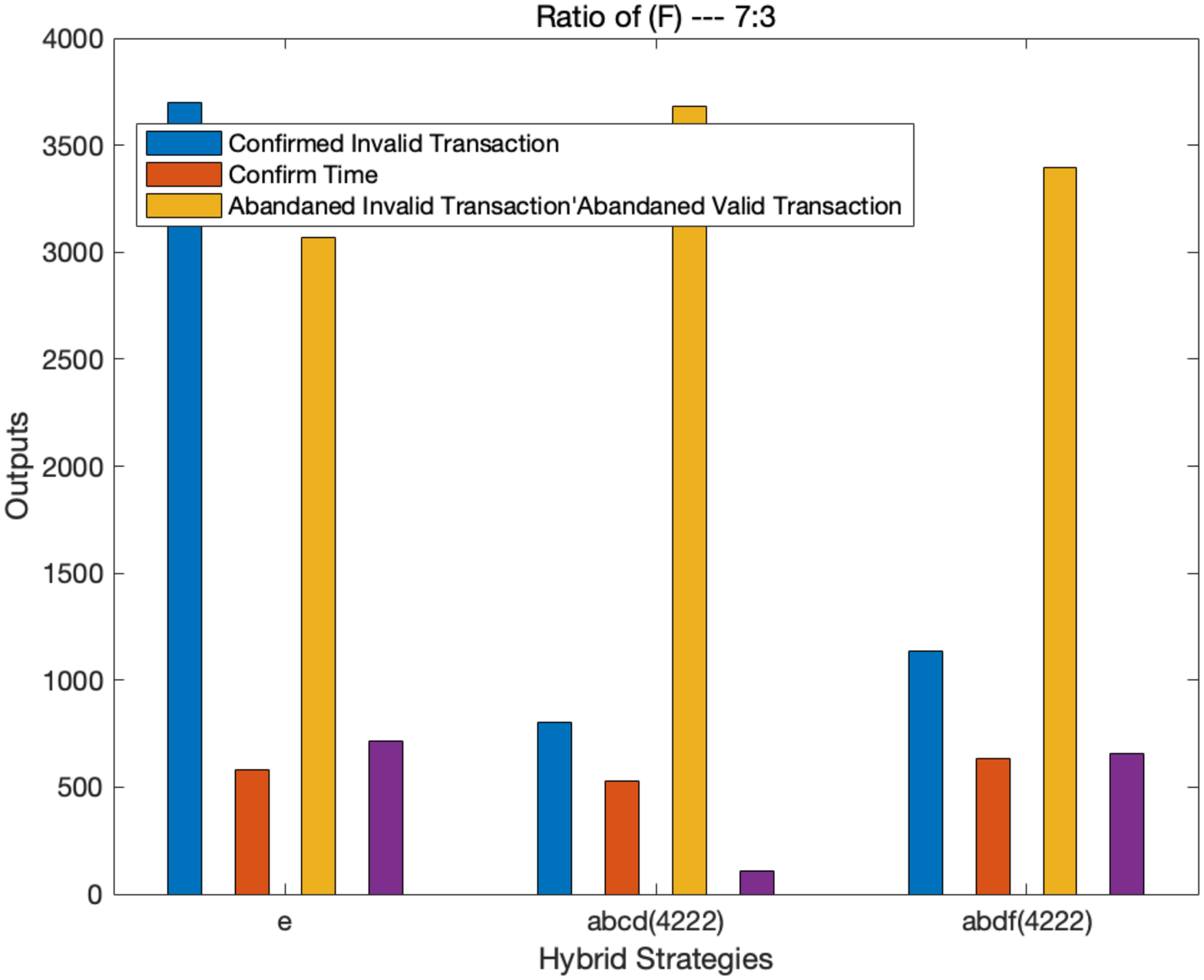}
\label{resulti}
\end{minipage}}
\subfigure[Result II]{
\begin{minipage}[b]{0.23\linewidth}
\includegraphics[width=1\linewidth]{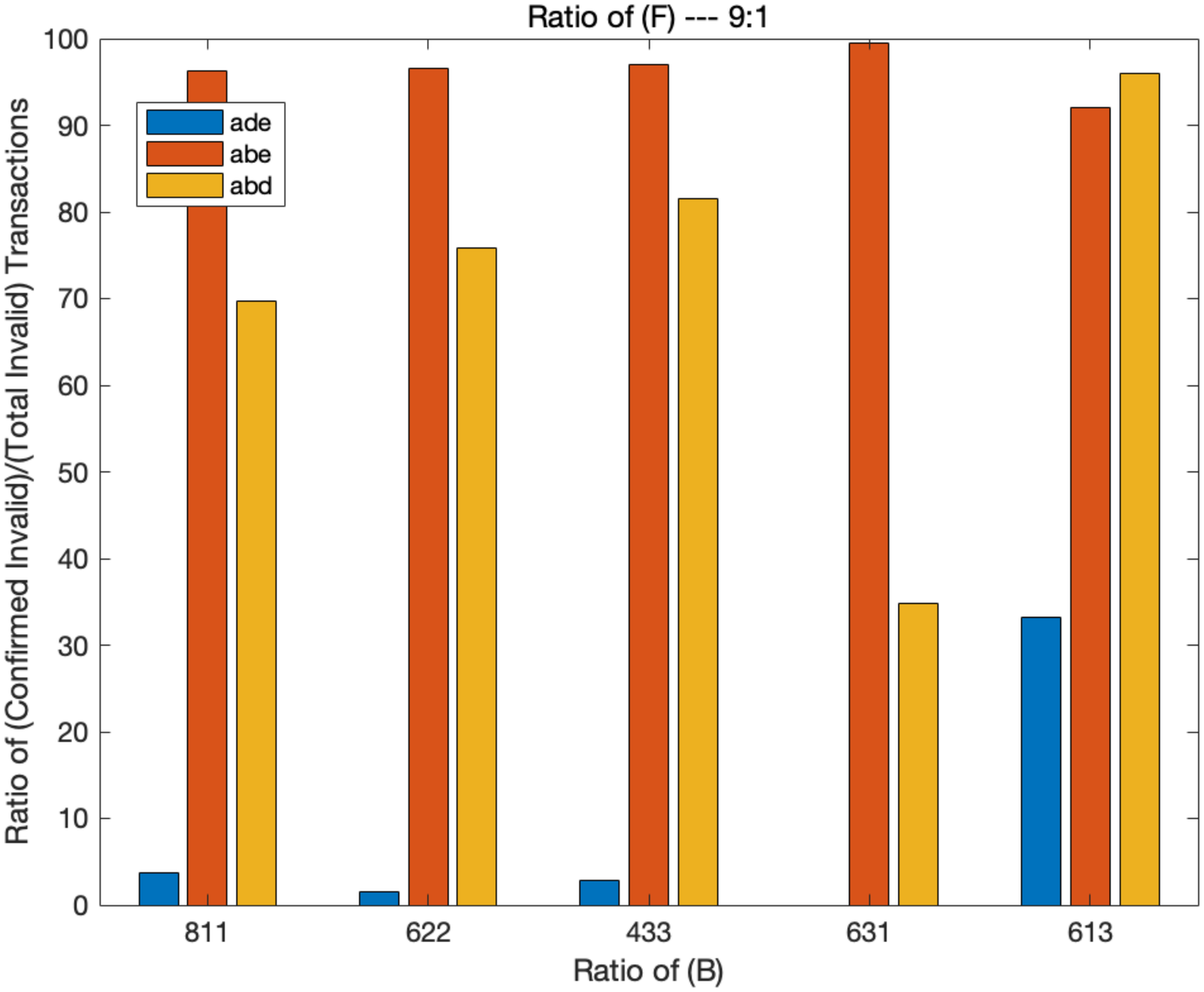}\vspace{4pt}
\includegraphics[width=1\linewidth]{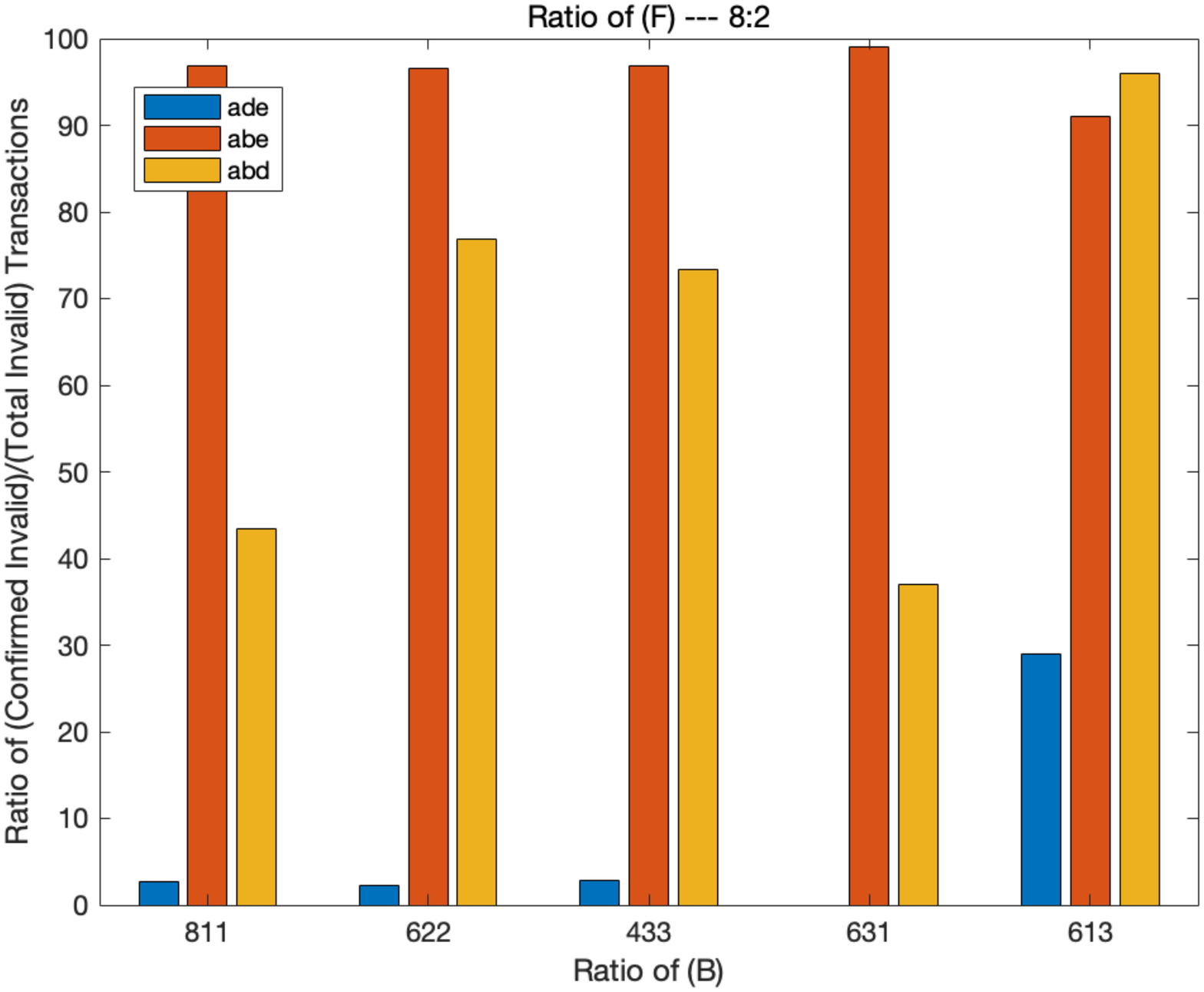}\vspace{4pt}
\includegraphics[width=1\linewidth]{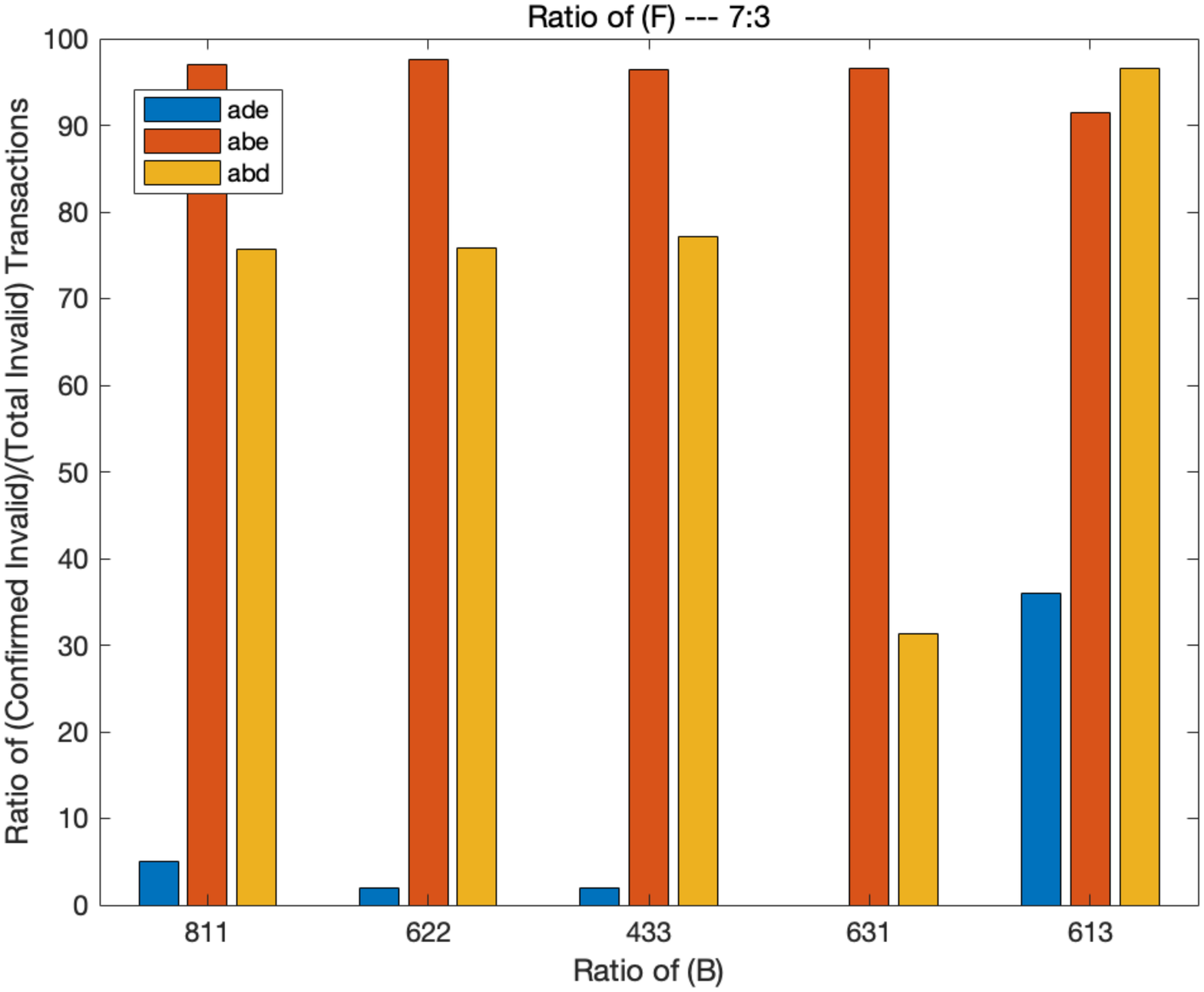}
\label{resultii}
\end{minipage}}
\subfigure[Result III]{
\begin{minipage}[b]{0.23\linewidth}
\includegraphics[width=1\linewidth]{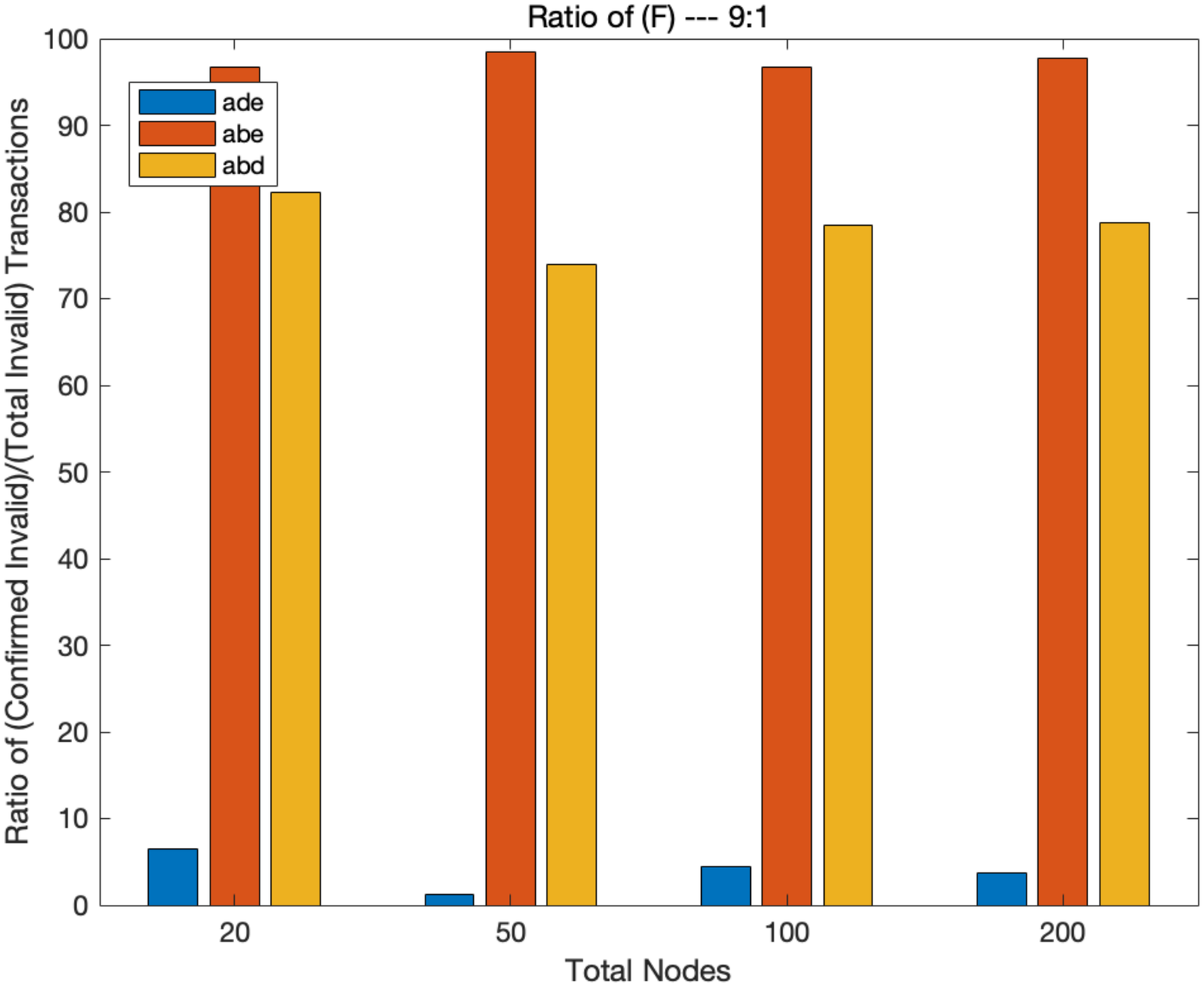}\vspace{4pt}
\includegraphics[width=1\linewidth]{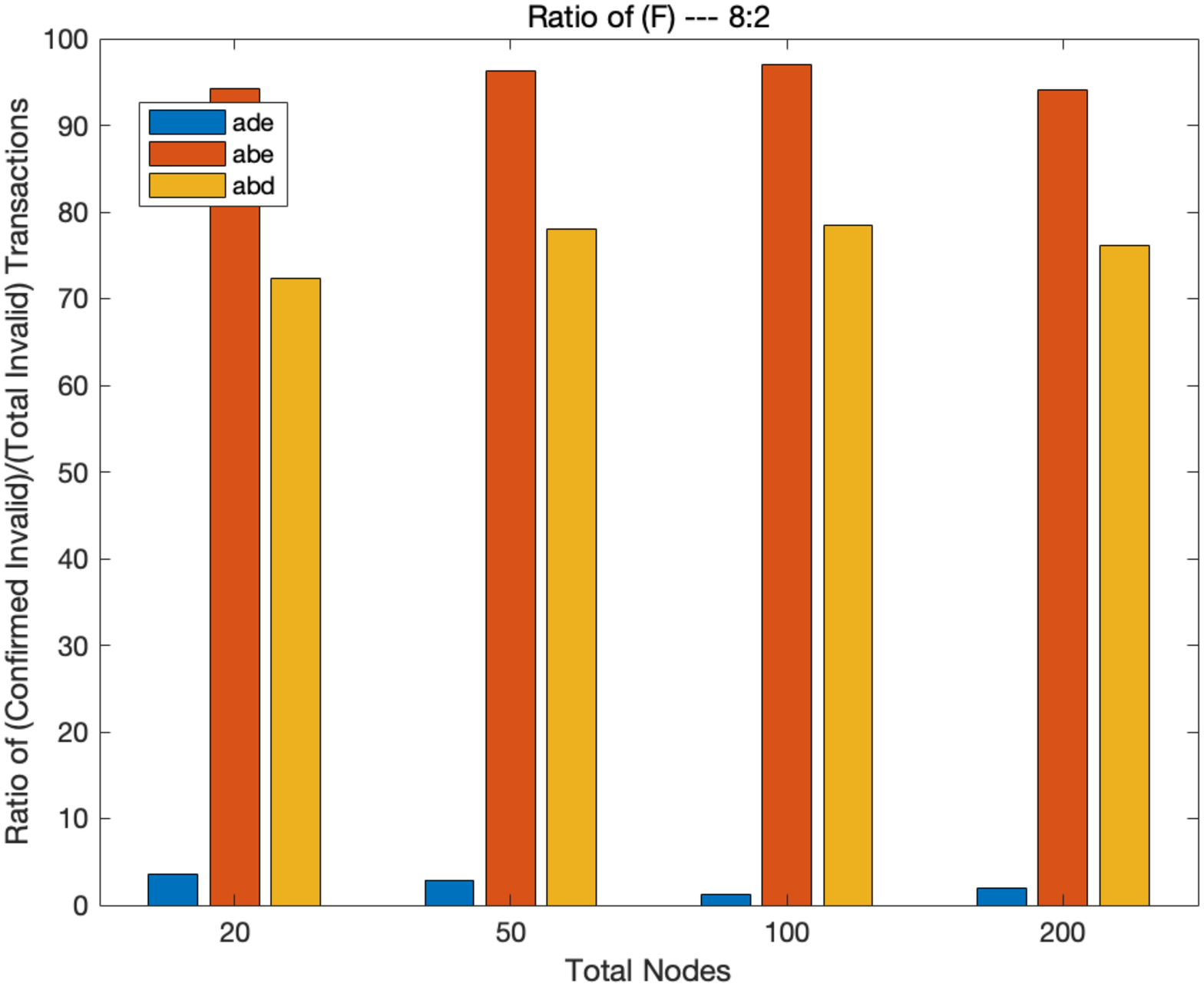}\vspace{4pt}
\includegraphics[width=1\linewidth]{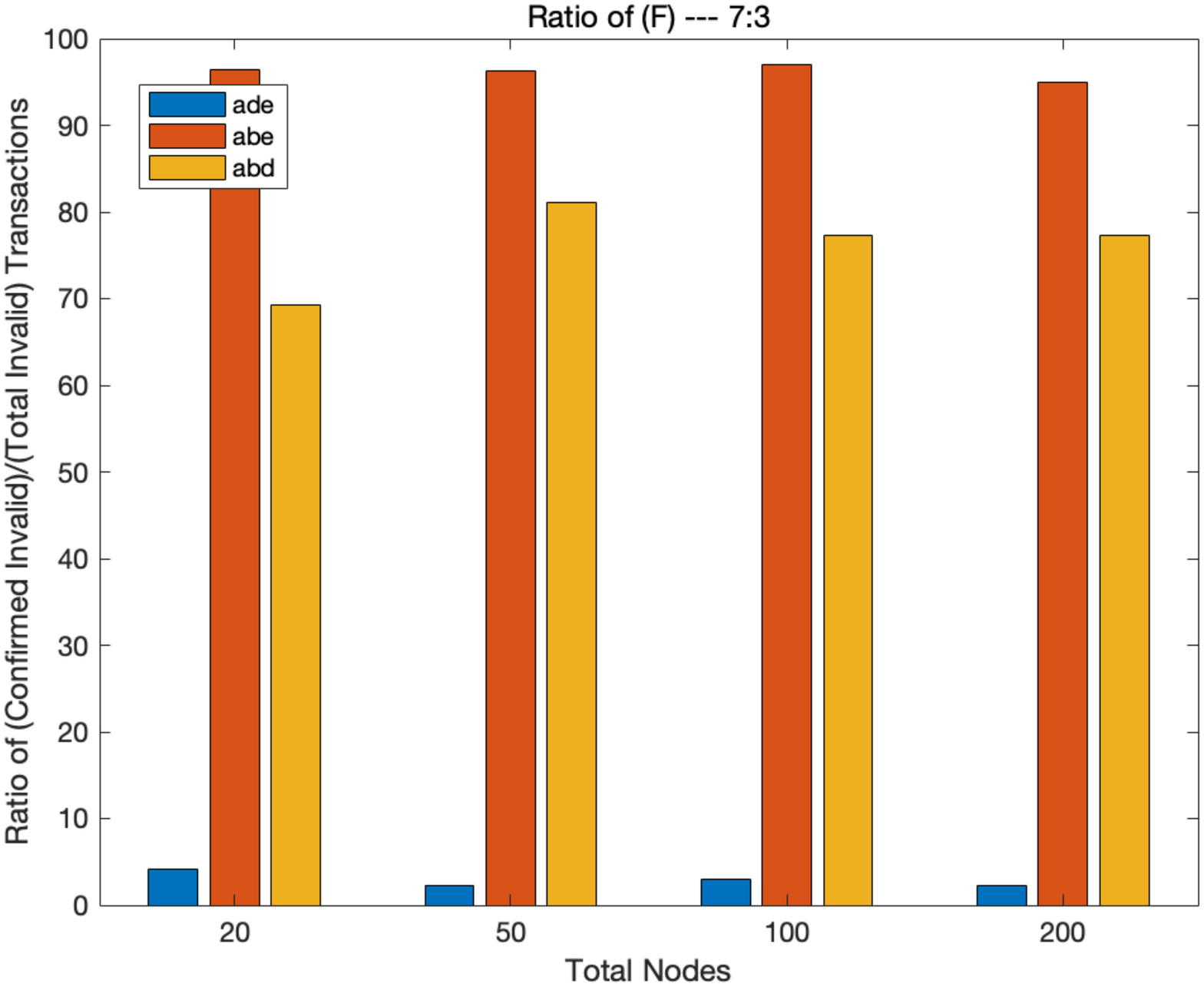}
\label{resultiii}
\end{minipage}}
\subfigure[Result IV]{
\begin{minipage}[b]{0.23\linewidth}
\includegraphics[width=1\linewidth]{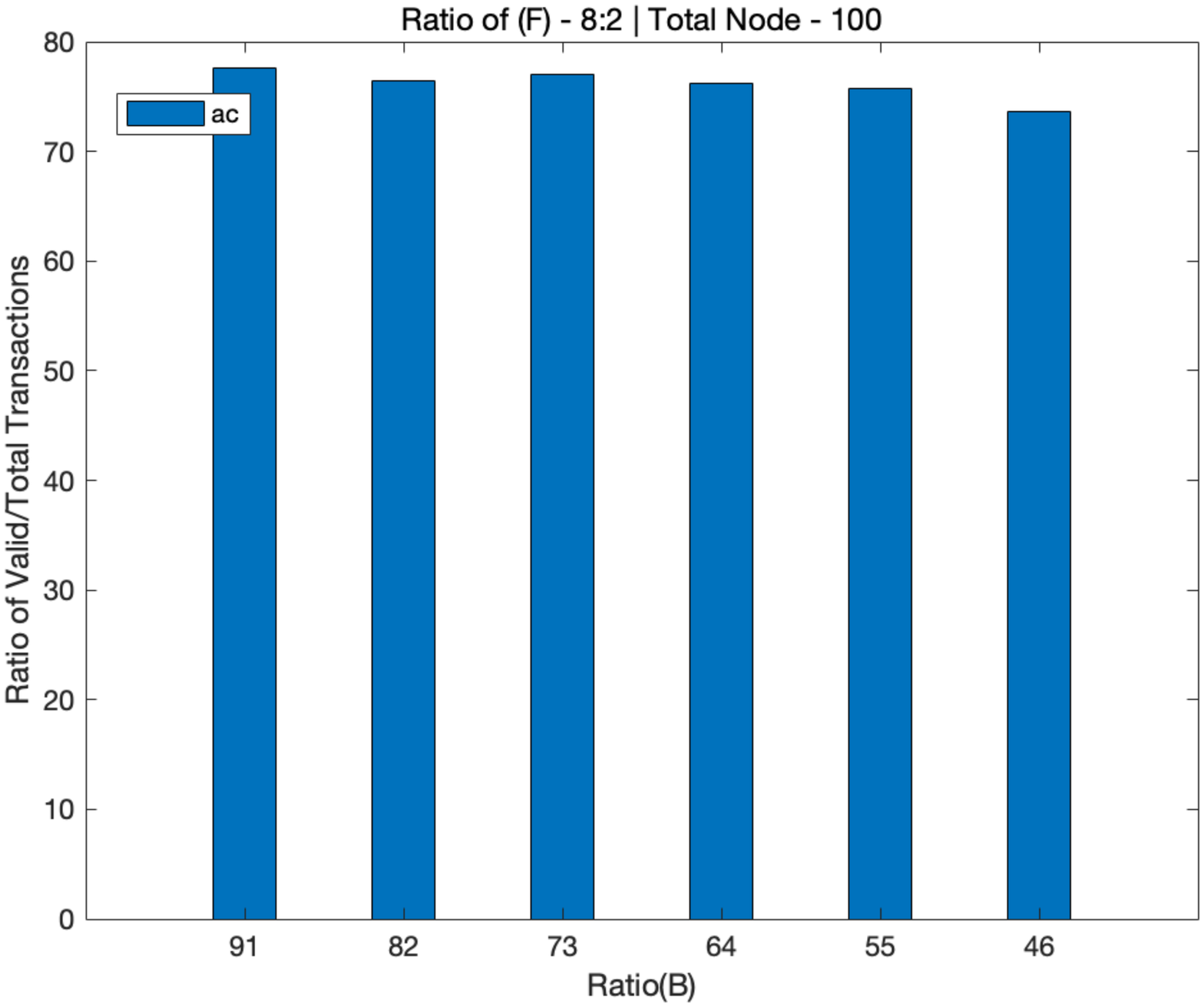}\vspace{4pt}
\includegraphics[width=1\linewidth]{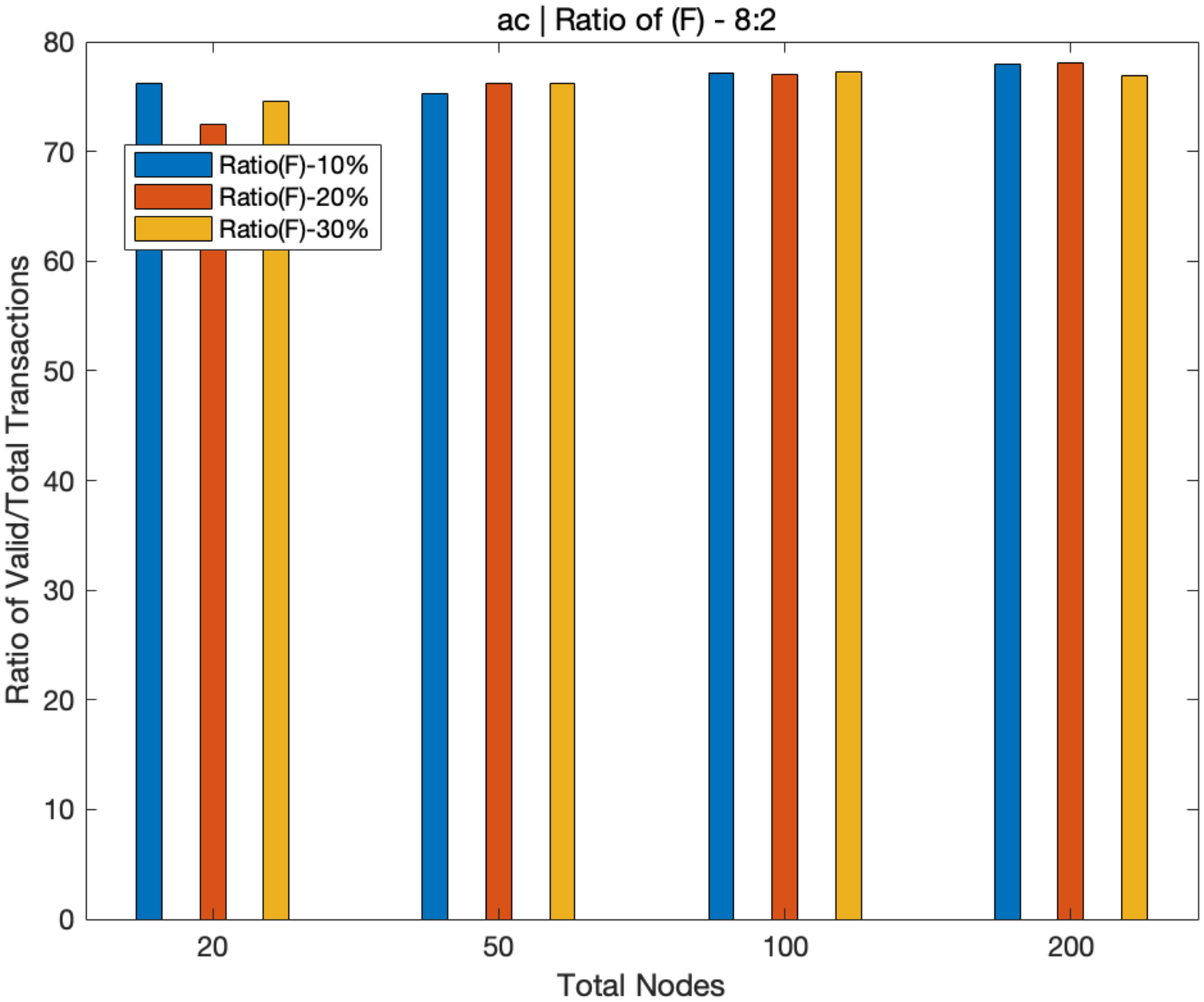}\vspace{4pt}
\includegraphics[width=1\linewidth]{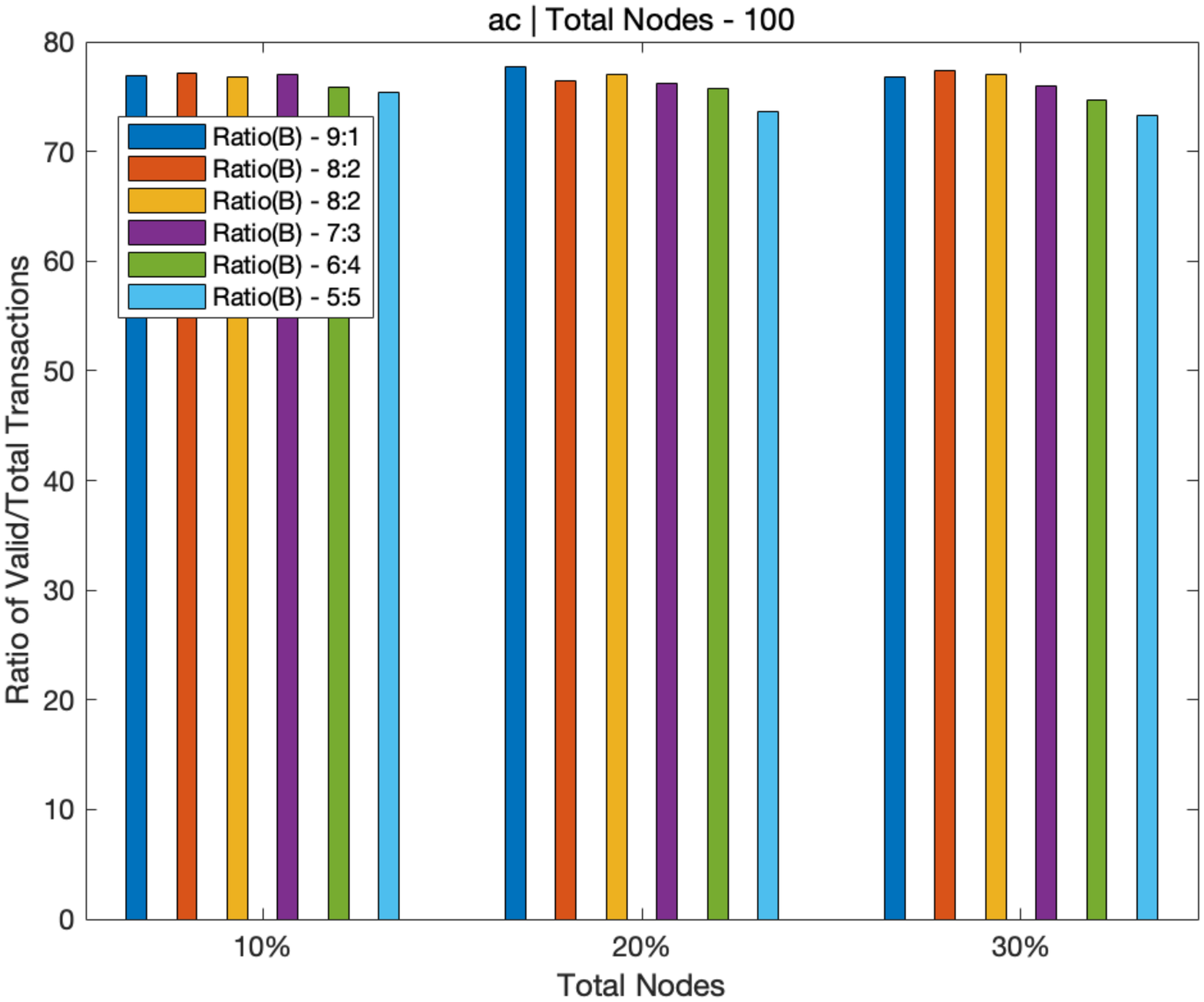}
\label{resultiv}
\end{minipage}}
\caption{Testing Results For Attack Strategies}
\label{pic-result}
\end{sidewaysfigure}

In the simulation IV, the test sets focus on the selfish strategy \textit{ac} where $\{ \mathbb{S} | \textit{ac}\}$. The first test initializes 100 total nodes, Ratio($\mathcal{F}$)= $\{ \mathcal{F} \,|\, 10\% \}$ and provides six Ratio($\mathcal{B}$)=$\{ \mathcal{B}\,|\, 91, 82, 73, 64, 55, 46 \} $. The second test set three Ratio($\mathcal{F}$)= $\{ \mathcal{F} \,|\, 10\%, 20\%, 30\%\}$, Ratio($\mathcal{B}$)= $\{ \mathcal{B} \,|\, 82\}$ and four sets of total nodes where $\{T\,|\, 20, 50, 100, 200 \}$. The third set initializes with 100 total nodes, Ratio($\mathcal{F}$)= $\{ \mathcal{F} \,|\, 10\%, 20\%, 30\%\}$ and Ratio($\{ \mathcal{B}\,|\, 91, 82, 73, 64, 55, 46 \} $) as the input parameters. The outputs are the ratio between valid transactions and total transactions which provides a direct and visualized relationship. As shown in the  \underline{\textit{Result IV} in Figure \ref{resultiv}}, we can find (1) the valid transactions maintain relatively stable under different Ratio($\mathcal{B}$). (2) The ratio is stable whenever the Ratio($\mathcal{F}$) increases or the total nodes increase. This means the ratio of valid transactions with total transactions vary slightly with the malicious nodes. Thus the number of malicious nodes has little influence. (3) The changes of Ratio($\mathcal{B}$) and Ratio($\mathcal{F}$) have slight influences on the results. All above-mentioned outcomes show that the selfish results only relates to the selfish behavior independent of its combination or strategy. Tangle maintains stable under the selfish behaviors.

\section{Discussions on Our Simulations}
\label{sec-discuss}

In this section, we provide the myths related to our simulations and analyses. Based on the myths below, we emphasize the importance and influence of our work in a plain and simple way.

\smallskip
\noindent\textit{Myth 1.Are Tangle-based blockchains important in DAG systems?}
DAG systems are emerging for several years, aims to improve the scalability by parallel processing. Tangle-based blockchains play the role of pioneer to inspire many other open-sourced projects. DAG projects based on blocks, such as Conflux \cite{li2018scaling}, partially change the original concept of scalability, since the final transactions need to be sequenced in a uniform order. At present, Tangle-based blockchain maximally inherent the property of scalability. The security analyses on such models are educational.

\smallskip
\noindent\textit{Myth 2.Does the simulation benefit for the real scenario?}
The simulation captures key features of real Tangle-based projects. The simulation cannot completely reflect real situations in a large network due to the design limitation. But it also provides an intuitive way to quantitatively analyze the security under different types of attacks. The results and trends demonstrate the potential vulnerabilities under multiple events in real scenarios, which would be a benefit for future design.

\smallskip
\noindent\textit{Myth 3.What the main factors of the attack effects?}
The attacks (DS, PS, HB) are sensitive to the Ratio($\mathcal{B}$), and the methods on how to make up a strategy are significantly influential to the attack effect. Preventing such effects needs to carefully consider all factors including binary actions, the ratio of behaviors, the ratio of malicious nodes, and the strategies under different combinations. However, the effect of selfish behavior is limited in a specified range, rather than attack strategies. The effect will appear when a strategy contains one or more selfish behaviors. Avoiding such effects requires to identify the selfish atomic behaviors from the bottom.

\smallskip
\noindent\textit{Myth 4.What could be improved learned from the simulation?} 
The simulation tests on Tangle-based attacks provide us several enlightening points. (1) The DAG can maintain stability in case of selfish behaviors no matter how it made up or how many selfish nodes exist. (2) The increasing malicious nodes will significantly increase the absolute number of transactions instead of probability since the successful attacks (the ratio of $\{$Confirmed Invalid Transaction$\}$/$\{$Total Invalid Transaction$\}$) maintains stable under different Ratio($\mathcal{F}$). (3) Tangle-based structures are sensitive to the binary actions in \textit{Layer0}, and the actions are deterministic for the final success.

\section{Conclusion}
\label{conclu}

Tangle, as one of the earliest DAG-based blockchain structures, provides an enlightening paradigm for peers. Various DAG-based blockchains with blockless structures are influenced by the concept of Tangle. We abstract the principles of Tangle-based blockchain from the basic actions (bottom) to the meshed graph network (top), including removing the structure of blocks, verifying multiple previous transactions, and configuring tip selection algorithms. Then, according to the above features, we simulate a simple but flexible network to construct and evaluate different types of attacks. Specifically, we define three actions as the basic benchmarks to construct our attack strategies layer by layer. Three types of attack strategies are provided, including parasite attack, double spending attack, and the hybrid attack. Each attack strategy is made up by multiple behaviors, where the behaviors are made up by different actions. We further evaluate these attacks in multi-dimensions with 12 sets of testing experiments. The evaluations cover the influence of both the binary selection of actions and the changeable parameters of configurations. Furthermore, we present a comprehensive discussion of the derivative question based on our constructions and evaluations. The results show the trends under different strategies and configurations. Our construction and evaluations provide an example for both attack and defense towards Tangle-based blockchains.

\smallskip
\noindent\textbf{Limitations and Future work.} The complicated testing goals and experimental results may confuse the readers. However, this paper mainly explains how to analyze a fresh new blockchain structure by building up a simulation model and progressively establishing on-top attack strategies. The comprehensive experiments are tested in multi-dimensions, where various aspects could be further studied. Detailed analyses on each single attack are not provided, such as whether 51$\%$ is enough for parasite attacks. Therefore, we will continue diving into more specific attacks in the future.

\bibliographystyle{splncs04}
\bibliography{bib}

\begin{thebibliography}{10}
\providecommand{\url}[1]{\texttt{#1}}
\providecommand{\urlprefix}{URL }
\providecommand{\doi}[1]{https://doi.org/#1}

\bibitem{nano}
Nano. https://nano.org/en/whitepaper  (2017)

\bibitem{dagwiki}
Dag. https://en.wikipedia.org/wiki/Directed\_acyclic\_graph  (2019)

\bibitem{baird2016swirlds}
Baird, L.: The swirlds hashgraph consensus algorithm: Fair, fast, byzantine
  fault tolerance. Swirlds Tech Reports SWIRLDS-TR-2016-01, Tech. Rep.  (2016)

\bibitem{boyen2016blockchain}
Boyen, X., Carr, C., Haines, T.: Blockchain-free cryptocurrencies: A framework
  for truly decentralised fast transactions. Cryptology ePrint Archive  (2016)

\bibitem{bu2019g}
Bu, G., G{\"u}rcan, {\"O}., Potop-Butucaru, M.: G-iota: Fair and confidence
  aware tangle. In: IEEE INFOCOM 2019-IEEE Conference on Computer
  Communications Workshops (INFOCOM WKSHPS). pp. 644--649. IEEE (2019)

\bibitem{CastroL99}
Castro, M., Liskov, B., et~al.: Practical byzantine fault tolerance. In: USENIX
  Symposium on Operating Systems Design and Implementation (OSDI). vol.~99, pp.
  173--186 (1999)

\bibitem{churyumov2016byteball}
Churyumov, A.: Byteball: A decentralized system for storage and transfer of
  value. URL https://byteball.org/Byteball.pdf  (2016)

\bibitem{cullen2020resilience}
Cullen, A., Ferraro, P., King, C., Shorten, R.: On the resilience of dag-based
  distributed ledgers in iot applications. IEEE Internet of Things Journal
  (2020)

\bibitem{cullen2019distributed}
Cullen, A., Ferraro, P., King, C., Shorten, R.: Distributed ledger technology
  for iot: Parasite chain attacks. arXiv preprint arXiv:1904.00996  (2019)

\bibitem{ferraro2018iota}
Ferraro, P., King, C., Shorten, R.: Iota-based directed acyclic graphs without
  orphans. arXiv preprint arXiv:1901.07302  (2018)

\bibitem{mam}
Foundation, I.: Masked authenticated messaging module.
  \url{https://github.com/iotaledger/MAM}  (2016)

\bibitem{qubic20}
Foundation, I.: Qubic: Accessed on june 01, 2020. \url{https://qubic.iota.org/}
   (2020)

\bibitem{gervais2016security}
Gervais, A., Karame, G.O., W{\"u}st, K., Glykantzis, V., Ritzdorf, H., Capkun,
  S.: On the security and performance of proof of work blockchains. In:
  Proceedings of the 2016 ACM SIGSAC conference on computer and communications
  security. pp. 3--16 (2016)

\bibitem{io2017eos}
IO, E.: Eos. io technical white paper. EOS. IO (accessed 18 December 2017)
  https://github. com/EOSIO/Documentation  (2017)

\bibitem{kusmierz2017first}
Kusmierz, B.: The first glance at the simulation of the tangle: discrete model.
  URL http://iota.org  (2017)

\bibitem{kusmierz2018extracting}
Kusmierz, B., Staupe, P., Gal, A.: Extracting tangle properties in continuous
  time via large-scale simulations. Tech. rep., working paper (2018)

\bibitem{lewenberg2015inclusive}
Lewenberg, Y., Sompolinsky, Y., Zohar, A.: Inclusive block chain protocols. In:
  International Conference on Financial Cryptography and Data Security (FC).
  pp. 528--547. Springer (2015)

\bibitem{li2018scaling}
Li, C., Li, P., Zhou, D., Xu, W., Long, F., Yao, A.: Scaling nakamoto consensus
  to thousands of transactions per second. arXiv preprint arXiv:1805.03870
  (2018)

\bibitem{nakamoto2008bitcoin}
Nakamoto, S.: Bitcoin: A peer-to-peer electronic cash system  (2008)

\bibitem{Nguyen2019StairDagCV}
Nguyen, Q.T., Cronje, A., Kong, M.G., Kampa, A., Samman, G.: Stairdag:
  Cross-dag validation for scalable bft consensus. ArXiv
  \textbf{abs/1908.11810} (2019)

\bibitem{popov2016tangle}
Popov, S.: The tangle. URL http://iota.org  (2016)

\bibitem{popov2017equilibria}
Popov, S., \textit{et al.}: Equilibria in the tangle. arXiv preprint
  arXiv:1712.05385  (2017)

\bibitem{sompolinskyserialization}
Sompolinsky, Y.Y., Zohar, A.: Serialization of proof-of-work events: Confirming
  transactions via recursive elections. IACR Cryptology ePrint Archive.

\bibitem{Sompolinsky2017SPECTRES}
Sompolinsky, Y., Lewenberg, Y., Zohar, A.: Spectre: Serialization of
  proof-of-work events : Confirming transactions via recursive elections
  (2017)

\bibitem{sompolinsky2013accelerating}
Sompolinsky, Y., Zohar, A.: Accelerating bitcoin’s transaction processing
  (2013)

\bibitem{sompolinsky2015secure}
Sompolinsky, Y., Zohar, A.: Secure high-rate transaction processing in bitcoin.
  In: International Conference on Financial Cryptography and Data Security
  (FC). pp. 507--527. Springer (2015)

\bibitem{qin2020security}
Wang, B., Wang, Q., Chen, S., Xiang, Y.: Security analysis on tangle-based
  blockchain through simulation. In: Australasian Conference on Information
  Security and Privacy (ACISP). pp. 653--663. Springer (2020)

\bibitem{wood2014ethereum}
Wood, G., et~al.: Ethereum: A secure decentralised generalised transaction
  ledger. Ethereum project yellow paper  \textbf{151}(2014),  1--32 (2014)

\end{thebibliography}

\newpage

\section*{Appendix A: More Details of IOTA}

\begin{figure}[H]
\centering
\includegraphics[width=0.8\linewidth]{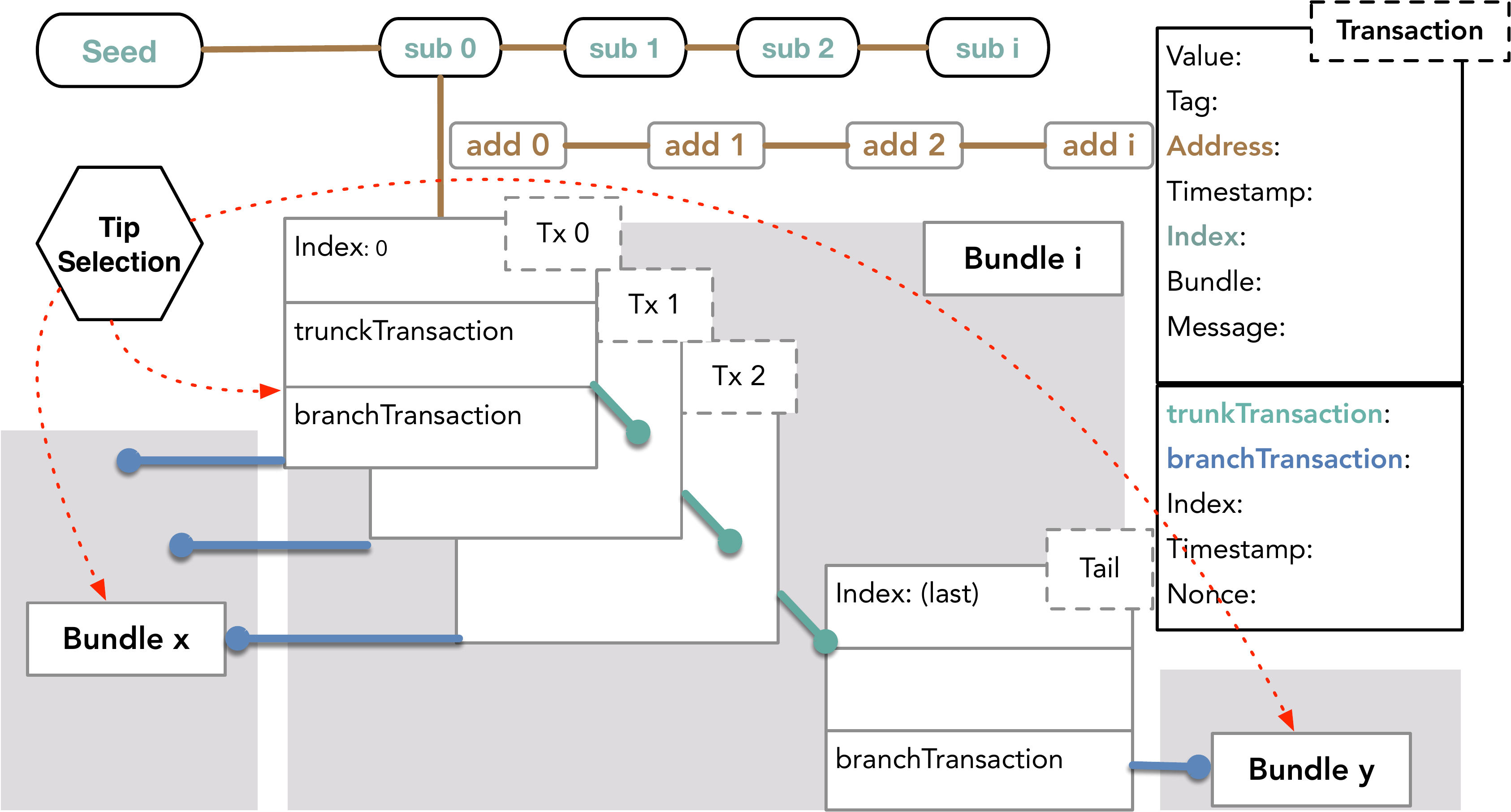}
\caption{Bundle Structure.}
\label{bundle}
\vspace{-10pt}
\end{figure}

\textbf{Bundle Structure.} \underline{Fig.\ref{bundle}} provides the bundle structure in Tangle. Several transactions are linked by the fields of \textit{trunckTransaction} and  \textit{branchTransaction}. This figure shows how to complete the token transferring.

\noindent\textbf{Extension of IOTA.} IOTA is a novel approach in improving the scalability of blockchain and eliminating the need for miners and transaction fees. IOTA comprises of the backbone protocol Tangle \cite{popov2016tangle}, and upper layer components such as MAM \cite{mam} for data communication and Qubic \cite{qubic20} for computations.  Tangle is the basic layer of IOTA to ensure its DAG-based transaction settlement and data integrity. It is essentially formed by individual transactions that are interlinked to each other. MAM, short for Masked Authenticated Messaging, is a second layer data communication protocol to provide the properties on integrity and privacy. MAM makes an encrypted data stream over Tangle regardless of the size or cost of device. Qubic, standing for quorum-based computation, is a protocol which specifies the solutions for oracle machines, outsourced computations, and smart contracts. It is used in distributed systems to gain consistent and fast computations with the least amount of costs possible.

\section*{Appendix B: Configurations on Goals}

The experiment goals describe the configurations on each testing set in detail. Goals I are tested in ($Set1,Set2,Set3$), while  Goals II in ($Set4,Set5,Set6$), Goals III in ($Set7,Set8,Set9$) and Goals IV in ($Set10,Set11,Set12$), respectively. Details are presented in the following \underline{Table \ref{tab-goals}}.

\begin{sidewaystable}[p]
  \centering
    \begin{tabular}[t]{cccc|cccc|cccc}
    \hline
    \hline
    \textit{Total Node}  &  \textit{Ratio($\mathcal{F}$)} & $\mathbb{S}$ &  \textit{Ratio($\mathcal{B}$)} & \textit{Total Node}  &  \textit{Ratio($\mathcal{F}$)} & $\mathbb{S}$ &  \textit{Ratio($\mathcal{B}$)} & \textit{Total Node}  &  \textit{Ratio($\mathcal{F}$)} & $\mathbb{S}$ &  \textit{Ratio($\mathcal{B}$)}   \\  
    \hline
    & Testing Set 1 & HB &&& Testing  Set 2 & HB && & Testing  Set 3 & HB \\  
    \cline{2-2} \cline{6-6} \cline{10-10} 
    100 & 20$\%$ & bd & 5:5 & 100 &  20$\%$ &  ade & 4:3:3 & 100 &  20$\%$  & e  & - \\  
    100 & 20$\%$ & be & 5:5 & 100 &  20$\%$ & abe & 4:3:3  & 100 &  20$\%$  & abcd &  4:2:2:2 \\  
    100 &  20$\%$ & - & - & 100 &  20$\%$ & abd & 4:3:3  & 100 &  20$\%$  & abdf & 4:2:2:2 \\  
   
    \hline
    \hline
    \textit{Total Node}  &  \textit{Ratio($\mathcal{F}$)} & $\mathbb{S}$ &  \textit{Ratio($\mathcal{B}$)} & \textit{Total Node}  &  \textit{Ratio($\mathcal{F}$)} & $\mathbb{S}$ &  \textit{Ratio($\mathcal{B}$)} & \textit{Total Node}  &  \textit{Ratio($\mathcal{F}$)} & $\mathbb{S}$ &  \textit{Ratio($\mathcal{B}$)}   \\  
    \hline
    & Testing Set 4 & PS/DS&&& Testing  Set 5 &PS/DS&& & Testing  Set 6 &PS/DS \\  
    \cline{2-2} \cline{6-6} \cline{10-10} 
    100 & 10$\%$ & *\footnote{``*'' represents the strategies of ade/abe/abd} & 8:1:1 & 100 &  20$\%$ & * & 8:1:1   & 100 &  30$\%$ & * & 8:1:1  \\  
    100 & 10$\%$ & * & 6:2:2 &  100 &  20$\%$ &  * & 6:2:2   & 100 &  30$\%$ &  * & 6:2:2  \\  
    100 &  10$\%$ & * & 4:3:3 &  100 &  20$\%$ &  * & 4:3:3   & 100 &  30$\%$ & * & 4:3:3  \\  
    100  &  10$\%$ & *  & 6:3:1 &  100 &  20$\%$ & *  & 6:3:1   & 100 &  30$\%$ & * & 6:3:1  \\  
    100 &  10$\%$ & * & 6:1:3 &  100 &  20$\%$ & * & 6:1:3  & 100 &  30$\%$ & * & 6:1:3  \\  
    \hline
    \hline
    \textit{Total Node}  & \textit{F} & \textit{$\mathbb{S}$} & \textit{Ratio($\mathcal{F}$)} & \textit{Total Node}   & \textit{F} & \textit{$\mathbb{S}$} & \textit{Ratio($\mathcal{F}$)}  & \textit{Total Node}  & \textit{F} & \textit{$\mathbb{S}$} & \textit{Ratio($\mathcal{F}$)}   \\  
    \hline
    & Testing Set 7 & &&& Testing  Set 8 &&& & Testing  Set 9 & \\  
    \cline{2-2} \cline{6-6} \cline{10-10} 
    20 &  2& * &10$\%$ & 20 &   4 &* & 20$\%$ & 20 &  6  &  * & 30$\%$\\  
    50 &  5 & * &10$\%$ & 50 & 10 & *  &  20$\%$ & 50 &   15  &  * & 30$\%$ \\  
    100 &  10 & * & 10$\%$  & 100 &   10 &* & 20$\%$ & 100 &  30  &  * & 30$\%$\\  
    200 &  20 & * & 10$\%$ & 200 &  40 & * & 20$\%$ & 200 &   60  &  * & 30$\%$ \\  
    \hline
    \hline
    \textit{Total Node}  & Ratio($\mathcal{F}$) & \textit{$\mathbb{S}$} & \textit{Ratio($\mathcal{B}$)} & \textit{Total Node}   &  \textit{Ratio($\mathcal{B}$)} & \multicolumn{2}{c|}{\textit{Ratio($\mathcal{F}$)} } & \textit{Total Node}  &  Ratio($\mathcal{B}$) &  \multicolumn{2}{c}{\textit{Ratio($\mathcal{F}$)} }   \\  
    \hline
    & Testing Set 10 & &&& Testing  Set 11 &&& & Testing  Set 12 & \\  
    \cline{2-2} \cline{6-6} \cline{10-10} 
    100 &  10$\%$ & ac  & 9:1 & 20 &  8:2  &  \multicolumn{2}{c|}{10$\%$, 20$\%$, 30$\%$}  & 100 &   9:1  &  \multicolumn{2}{c}{10$\%$, 20$\%$, 30$\%$}  \\  
    100 &  10$\%$ & ac  & 8:2 & 50 &  8:2  &  \multicolumn{2}{c|}{10$\%$, 20$\%$, 30$\%$} & 100 &    8:2  &  \multicolumn{2}{c}{10$\%$, 20$\%$, 30$\%$} \\  
    100 &  10$\%$ & ac  & 7:3 & 100 &  8:2  &  \multicolumn{2}{c|}{10$\%$, 20$\%$, 30$\%$} & 100 &   7:3  &   \multicolumn{2}{c}{10$\%$, 20$\%$, 30$\%$}  \\  
    100 &  10$\%$ & ac  & 6:4 & 200 &  8:2 &  \multicolumn{2}{c|}{10$\%$, 20$\%$, 30$\%$} & 100 &    6:4  &  \multicolumn{2}{c}{10$\%$, 20$\%$, 30$\%$}  \\  
    100 &  10$\%$ & ac  & 5:5 & - &  - &  \multicolumn{2}{c|}{10$\%$, 20$\%$, 30$\%$} & 100 &    5:5  &   \multicolumn{2}{c}{10$\%$, 20$\%$, 30$\%$}  \\ 
    100 &  10$\%$ & ac  & 4:6 & - &  - &  \multicolumn{2}{c|}{10$\%$, 20$\%$, 30$\%$} & 100 &    4:6  &   \multicolumn{2}{c}{10$\%$, 20$\%$, 30$\%$}  \\
    \hline
    \hline
    \end{tabular}
    \caption{Configurations on Goals}
    \label{tab-goals}
\end{sidewaystable}

\clearpage
\section*{Appendix C: Example Code}
The tip selection mechanism is closely related to the cumulative weight of transactions.  Here, we provide the example codes of weight calculation, named $\mathsf{addweight}$.

\begin{figure}[H]
\centering
\includegraphics[width=\linewidth]{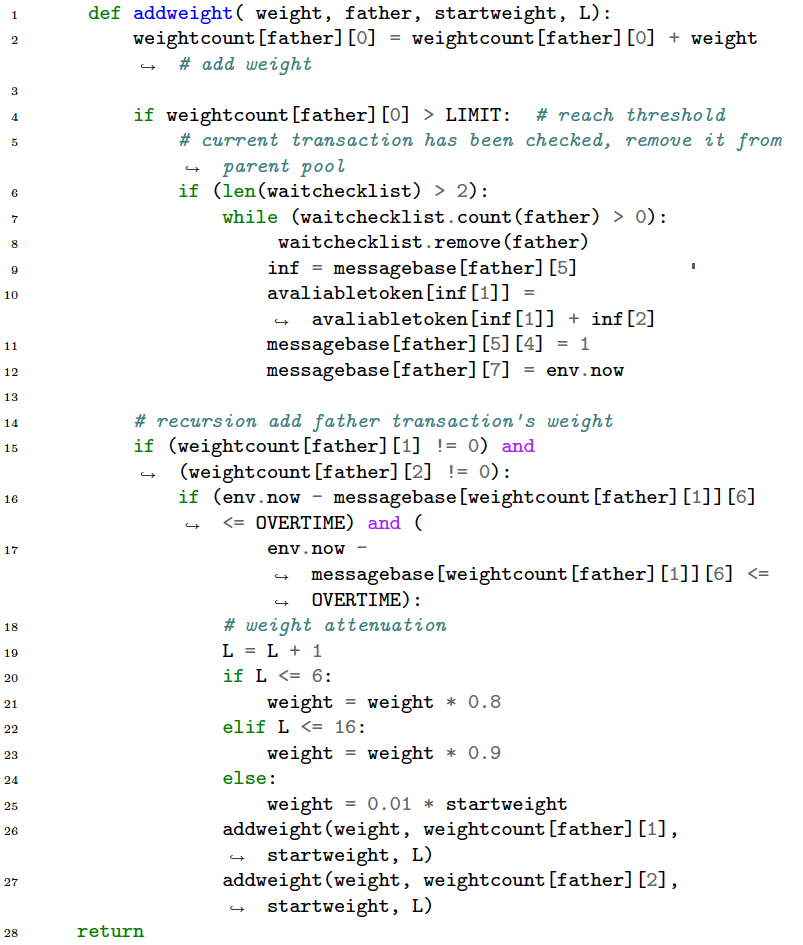}
\end{figure}

\section*{Appendix D: Test Results}



\end{document}